\documentclass[journal,comsoc,onecolumn,10pt]{IEEEtran}
\usepackage{cite}
\usepackage{amsmath,amssymb,amsfonts}
\usepackage{algorithmic}
\usepackage{graphicx}
\usepackage{textcomp}
\usepackage{wrapfig,colortbl}
\usepackage{caption}
\usepackage{comment}
\usepackage{xcolor}
\usepackage{color, soul}
\usepackage{tabularx}
\usepackage{array}
\usepackage{tabularray}
\usepackage{soul}
\usepackage{multirow}
\usepackage{subfig,graphicx}
\usepackage{layouts}
\usepackage{epstopdf}

\graphicspath {{./Introduction/}{./Experimental_setup/}{./Random_noise/}{./Coherent_noise/}{./Generic_scheme/}{./Simualtion_validation/}{./Experimental_validation/}{./Reliability_estimation/}}

\begin{document}
\title{Efficient generation of realistic guided wave signals for reliability estimation}
\author{Panpan Xu, Robin Jones, Georgios Sarris, and Peter Huthwaite  \IEEEmembership{}
\thanks{This work was supported by GW4SHM, EU Horizon 2020 H2020-MSCA-ITN-2019, Grant No. 860104.}
\thanks{Panpan Xu, Georgios Sarris, and Peter Huthwaite are with the Non-Destructive Evaluation Group, Mechanical Engineering Department, Imperial College London, SW7 2AZ London, UK. (e-mail: p.xu20@imperial.ac.uk, p.huthwaite@imperial.ac.uk).}
\thanks{Robin Jones is with Guided Ultrasonics Ltd., Brentford, TW8 8HQ, UK (e-mail: robin.jones@guided-ultrasonics.com).}
}

\thispagestyle{empty} 
\clearpage
\pagenumbering{arabic} 
\maketitle

\begin{abstract}

Across non-destructive testing (NDT) and structural health monitoring (SHM), accurate knowledge of the systems' reliability for detecting defects, such as Probability of Detection (POD) analysis is essential to enabling widespread adoption. Traditionally this relies on access to extensive experimental data to cover all critical areas of the parametric space, which becomes expensive, and heavily undermines the benefit such systems bring. In response to these challenges, reliability estimation based on numerical simulation emerges as a practical solution, offering enhanced efficiency and cost-effectiveness. Nevertheless, precise reliability estimation demands that the simulated data faithfully represents the real-world performance. In this context, a numerical framework tailored to generate realistic signals for reliability estimation purposes is presented here, focusing on the application of guided wave SHM for pipe monitoring. It specifically incorporates key characteristics of real signals: random noise and coherent noise caused by the imbalance in transducer performance within guided wave monitoring systems. The effectiveness of our proposed methodology is demonstrated through a comprehensive comparative analysis between simulation-generated signals and experimental signals both individually and statistically. Furthermore, to assess the reliability of a guided wave system in terms of the inspection range for pipe monitoring, a series of POD analyses using simulation-generated data were conducted. The comparison of POD curves derived from ideal and realistic simulation data underscores the necessity of considering coherent noise for accurate POD curve calculations. Moreover, the POD analysis based on realistic simulation-generated data provides a quantitative estimation of the inspection range with more details compared to the current industry practice. Our presented framework offers a pioneering approach to generate realistic guided wave signals, thereby facilitating the practical assessment of the reliability of guided wave monitoring systems. This advancement also has the potential to effectively address challenges related to data scarcity in broader applications requiring high-fidelity data, such as the training of machine learning models for damage identification from complex signals for all aspects of ultrasonic inspections with both guided and bulk waves.

\end{abstract}


\section{Introduction}
\label{sec:introduction}

\IEEEPARstart  {U}{ltrasonic} inspection has broad impact in industry for being a safe, subsurface modality. Within this, guided wave testing is a highly appealing non-destructive testing (NDT) method, particularly for inspecting large structural areas, owing to its ability to cover significant portions of a structure from a single transducer location. Its applications in the inspection of pipelines within the oil and gas industry have been well-established\cite{cawley2003practical,Peter10.1115/1.2789107}. Recent advancements in permanently installed transducers have made it increasingly attractive to transition from periodic NDT inspections to Structural Health Monitoring (SHM), which promises enhanced reliability and a reduction in the operational costs typically associated with regular inspections\cite{croxford2007strategies,cawley2012guided,Mariani:2023,giurgiutiu2010guided}. In addition, fully automating the monitoring can significantly reduce the operational risks for personnel and operators. Despite the progress made in advancing guided wave SHM from laboratory experimentation to practical industrial applications\cite{Vogt:2021}, the assessment of reliability remains a critical step for its widespread deployment in industry\cite{Cawley:2019,cawley2021development,mesnil2017guided}. Conventional reliability estimation methods, such as Probability of Detection (POD) analysis\cite{hdbk2009nondestructive}, rely on a statistically significant number of experimental trials, which involve manufacturing representative samples with various flaw sizes and measuring them under realistic conditions. The challenge lies in the expense and time intensiveness of experimental approaches to establish reliability estimates. \par
In response to this challenge, the concept of Model-Assisted Probability of Detection (MAPOD) was introduced\cite{ModelAssistedPODWorkingGroup,thompson2009recent}, aiming to substitute empirical experiments with physics-based theoretical or numerical models for reliability estimation, thereby effectively reducing the cost and time associated with sample fabrication and experimentation. Since its introduction, the MAPOD method has garnered considerable attention and has been extensively explored in the realms of NDT\cite{du2019multifidelity,yilmaz2022model} and SHM\cite{Pierre2023,falcetelli2023model}. Early employment of MAPOD has been applied in the POD estimation of eddy current inspection\cite{aldrin2009model} and ultrasonic measurement\cite{smith2007model}. There have also been several implementations of MAPOD in ultrasonic guided wave inspection. Moriot’s studies\cite{moriot2018model} used a simplistic finite element (FE) model to calculate the POD and Possibility of Location (POL) curves for assessing the performance of a guided wave-based SHM imaging system. Howard and Cegla\cite{howard2017probability,howard2017detectability} applied a MAPOD approach to evaluate the capability of short-range circumferential guided waves in detecting corrosion damage in the pipe. More recently, a simulation tool based on the CIVA simulation platform has been developed for the calculation of POD curves, demonstrating several successful applications in aerospace inspection\cite{foucher2018new}. While numerical tools offer the flexibility to generate data with varying configurations compared to theoretical models, the challenge remains in fully capturing the effects of practical measurement conditions.\par
The accuracy of the reliability estimation relies on the fidelity of the model-generated data, which is required to accurately capture the complexities of real experimental data, which are inherently noisy. Efforts have been made in the literature to make the simulation-generated data more realistic. For example, Gaussian white noise is commonly added to simulation data to generate noisy data\cite{khurjekar2019accounting}. However, actual experimental signals not only exhibit random noise but also coherent noise due to imperfect guided wave excitation conditions. In Mariani’s study\cite{mariani2020compensation}, variations in the amplitude of the excitation signal, representing the imbalance of transducer performance, were used to represent the coherent noise in the fundamental torsional mode, known as the T(0,1) mode, signals. However, phase variations also exist alongside amplitude variations, again due to inherent transducer imbalance\cite{herdovics2019compensation}, and the performance of the transducers is also influenced by varying environmental conditions, thereby influencing the noise levels in the signal. Liu et al.\cite{liu2017efficient} proposed a hybrid data generation framework to represent the environmental effects, which superposes experimental data collected under different environmental conditions on an undamaged structure with artificial damage signals. However, the influence of environmental conditions on the artificial damage responses themselves and the distortion of coherent noise caused by mode conversions and reverberations from the defect were not taken into account, making the synthetic signals less representative of the real signals. More recently, generative adversarial networks (GAN) based models have emerged as a promising method for data generation\cite{luleci2022generative}, yet the training process requires a substantial amount of data, and the generated signals lack practical explainability.\par
Despite the promising potential of model-generated data in estimating the reliability of guided wave monitoring systems, a high-fidelity model for generating guided wave signals remains underdeveloped. Therefore, this paper aims to establish, for the first time, a comprehensive numerical framework for generating realistic guided wave signals, informed by a thorough understanding of noise characteristics and contributors, with a focus on guided wave SHM for pipe monitoring. The simulation-generated realistic guided wave signals are then used to assess the performance of a guided wave pipe monitoring system in terms of the inspection range via POD analysis. Our presented framework offers a pioneering approach to generate realistic guided wave signals, and in addition to the benefits for assessing the reliability of guided wave monitoring systems, this step-change in capability will be vital for providing data for training machine learning models for damage identification from complex guided wave signals, as well as for training and assessment of human operators. While the focus in this paper is specifically on guided wave pipe monitoring applications, the techniques developed here are also applicable, with few changes, to other areas of inspection and monitoring across a range of modalities.\par
The subsequent sections are structured as follows: Section \ref{Real guided wave signals and noise} presents a comprehensive analysis of noise, encompassing coherent noise and random noise, based on experimental data. Section \ref{Methodology} outlines the proposed numerical framework for generating realistic guided wave signals, including simulation and experimental validations, along with an error analysis. In Section \ref{Reliability estimation}, a quantitative estimation method for determining the inspection range of a guided wave SHM system for pipe monitoring is detailed, based on simulation data generated from the proposed numerical framework in this study. Finally, the paper concludes with a summary in Section \ref{Conclusion}.\par

\section{Real guided wave signals and noise}\label{Real guided wave signals and noise}

Standard metallic pipe monitoring based on guided waves operates by transmitting a single guided wave mode and subsequently identifying the scattering waves arising from any irregularities within the pipe. A well-established approach for pipe monitoring is based on the T(0,1) mode, with frequencies typically under 150 kHz. This wave mode is non-dispersive, simplifying the interpretation of guided wave signals. Nevertheless, the guided wave signals in real pipes are nearly always contaminated by noise, which potentially obscures scattered waves from defects, masking the desired signals and complicating interpretation. Therefore, the performance of a guided wave monitoring system relies on the system's capability to differentiate between scattering waves originating from structural discontinuities and the underlying noise.\par
\begin{figure*}[hbt!] 
    \centering
    \includegraphics{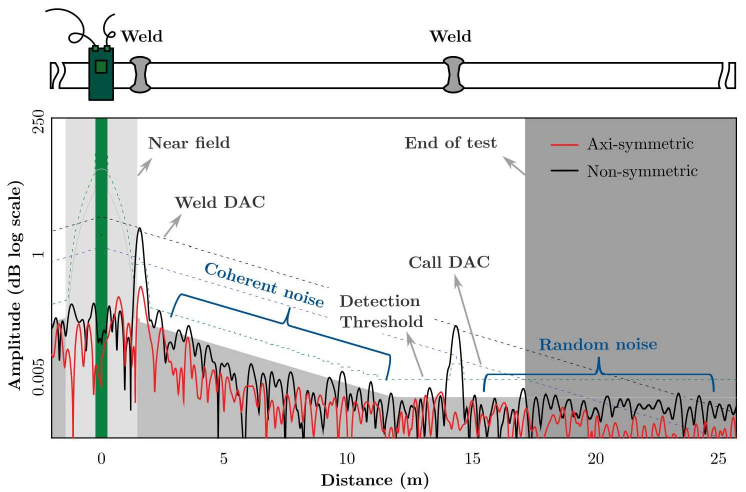}
    \caption{A typical guided wave signal (on the logarithmic scale) acquired by Guided Ultrasonics Ltd. on a pipe with two weld features.}
    \label{Figure 1}
\end{figure*}

Figure \ref{Figure 1} illustrates a typical guided wave signal acquired and processed by a commercial guided wave monitoring system from Guided Ultrasonics Ltd. According to the ASTM standard practice for guided wave testing of steel pipework\cite{e07_committee_practice_nodate}, Distance Amplitude Correction (DAC) curves are used to assess the attenuation and amplitude reduction over wave propagation distances, aiding in the assessment of damage extent, particularly in terms of pipe's cross-sectional change (CSC). In Fig. \ref{Figure 1}, DAC curves (dotted lines) were generated during post-processing using the Absolute Calibration method\cite{vogt2022multiple}, based on known pipe features such as welds. The commonly used DAC curves include Flange DAC, Weld DAC, and Call DAC. Flange DAC illustrates the anticipated amplitude reflected from a feature with an approximate 100\% reflection coefficient, while Weld DAC demonstrates the anticipated amplitude reflected from a pipe girth weld, typically presenting 20\% to 25\% CSC. The Call DAC level is generally established at roughly 6\% CSC, representing the system's sensitivity to detect the defect. Additionally, the noise level is illustrated as the dark grey area in Fig. \ref{Figure 1}, and the Detection Threshold is typically set at 6 dB above the noise level to avoid false alarms in defect detection.\par
In Fig. \ref{Figure 1}, the guided wave signals are displayed on the logarithmic scale and present both the axi-symmetric T(0,1) mode (shown in black) and the non-symmetric flexural mode (shown in red). Non-symmetric flexural modes arise from either direct excitation through non-axisymmetric transduction or mode conversion, which happens when the axi-symmetric T(0,1) incident wave interacts with non-symmetric features like simple supports. Alongside the reflections from the pipe weld features, discernible noise is also present. The noise within a guided wave signal consists of both random noise and coherent noise\cite{evans2010reliability}. Random noise originates from the electrical interference of the measurement system or environmental vibrations and can be reduced through averaging multiple signals together; because of the stochastic randomness, the different realisations of random noise sum incoherently and are reduced in amplitude, while the features which remain constant between the measurements, such as defect responses, are enhanced. Coherent noise, on the other hand, results from physical wave behaviour, such as the propagation of unwanted modes\cite{mariani2019location} or scattering from the pipe's rough surface, and is constant between repeat measurements. Notably, random noise maintains a consistent amplitude with respect to distance from the transducer ring, whereas the amplitude of coherent noise diminishes with the wave propagation distance, manifesting a linear decay on the logarithmic scale due to damping and beam spread effect.\par

In this section, a comprehensive analysis of the noise signal is performed based on the guided wave signals collected from an experimental setup within controlled laboratory environment at a room temperature of around 23 °C and humidity of around 25\%. This analysis serves as the basis for the subsequent modelling of noise in the simulations for generating realistic guided wave signals. 

\subsection{Experimental setup}
\begin{figure}[!t]
\centerline{\includegraphics[width=0.6\columnwidth]{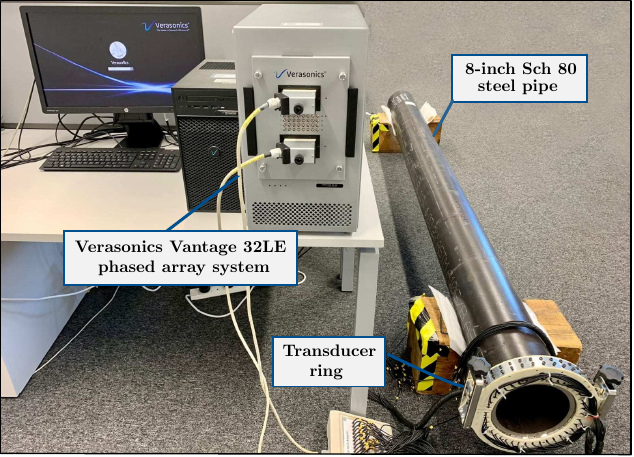}}
\caption{Experimental setup.}
\label{Figure 2}
\end{figure}

In the experiments, two API (American Petroleum Institute) standard seamless carbon steel pipes of different sizes, Pipe 1 and Pipe 2, as listed in Table \ref{Table 1}, were tested. The material properties of the pipes were referenced from a public engineering material database\cite{Material} but were calibrated based on the wave velocities measured experimentally to account for variations in material properties due to uncertainty in the exact processing parameters during manufacturing. The geometric and material information of the pipes is detailed in Table \ref{Table 1}, and the experimental setup is illustrated in Fig. \ref{Figure 2}. \par
\begin{figure}[!t] 
\centerline{\includegraphics{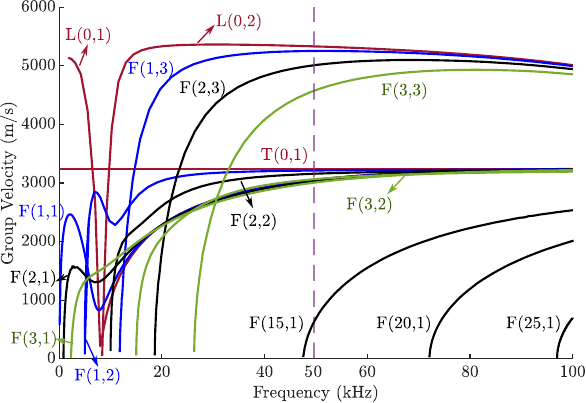}}
\caption{The group velocity dispersion curves of Pipe 2, an 8-inch Schedule 80 steel pipe.}
\label{Figure 7}
\end{figure}

For the excitation and reception of the T(0,1) wave mode, a transducer ring with 40 equally-spaced transducer elements, which apply shear forces in the circumferential direction, was installed at one of the ends of the pipe. This positioning allowed for the longest inspection range and avoided the need for directional control. The excitation signal was a 5-cycle Hann-windowed toneburst signal with a central frequency of 50 kHz. The high number of transducer elements around the pipe circumference theoretically ensured the suppression of all flexural modes within the frequency of interest, as shown in the disperse curve in Fig. \ref{Figure 7}. The dispersion curve was calculated based on Pipe 2 using Disperse\cite{pavlakovic1997disperse}. The small interelement space, approximately 1/4 wavelength in size, ensured the formation of effective plane waves without the occurrence of grating lobes\cite{wilcox2003omni}. A Verasonics Vantage 32LE phased array measurement system was employed for signal excitation and reception, with a sampling frequency of 2 MHz. The collected experimental signals were filtered using a bandpass filter, with a lower frequency limit of 36 kHz and an upper frequency limit of 64 kHz, to eliminate extraneous noise beyond the frequency range of interest. The bandpass filter limits were determined through trials, adjusting the limits and assessing the noise level compared to the desired response from the system.\par

The T(0,1) signals were collected using the common source method (CSM)\cite{Davies10.1063/1.2184522}. In the CSM arrangement, all transmitters were fired simultaneously to allow a plane wave to propagate axially. Simultaneously, signals from all receivers were captured and summed to obtain the T(0,1) signal.  \par

\newcolumntype{P}[1]{>{\centering\arraybackslash}p{#1}}
\newcolumntype{M}[1]{>{\centering\arraybackslash}m{#1}}
\begin{table*}[!h]
\renewcommand{\arraystretch}{2}
\caption{Geometric and material information of the instrumented pipes used in the experiments.}

\label{Table 1}
\begin{center}
\begin{tabular}{M{0.9cm}|M{1.2cm}|M{1.0cm}|M{1.2cm}|M{1.4cm}|M{1.2cm}|M{1.2cm}|M{1.2cm}|M{1.2cm}|M{1.2cm}|M{2cm}}

   \hline\hline
\rule{0pt}{15pt}\textbf{} & \multicolumn{5}{ c |}{\textbf{Geometric}} & \multirow{3.5}{*}{\textbf{Material}}   & \multicolumn{4}{ c }{\textbf{Material Properties}}\\ 
\cline{2-6}\cline{8-11}
\rule{0pt}{18pt} &\textbf{Nominal Size (inches)}&\textbf{Schedule}  & \textbf{Outside Diameter (mm)} & \textbf{Wall Thickness (mm)} & \textbf{Length (mm)} & & \textbf{Density (kg/m$^3$)} & \textbf{Young's Modulus (GPa)} & \textbf{Poisson's Ratio}& \textbf{Mass Proportional Damping Coefficient (1/second)} \\
\hline

\rule{0pt}{15pt} Pipe 1 & 8 & 60 & 219.1 & 10.31 &2370 &P265GH &7850&212.56&0.294&0\\
\hline
\rule{0pt}{15pt} Pipe 2& 8 & 80 & 219.1 & 13.3 & 3000&P265GH& 7850&210.86&0.285&0\\ 

   \hline\hline
\end{tabular}
\end{center}
\end{table*}

\subsection{Noise collection}
\begin{figure}
\centering  
\subfloat[]{\label{Figure 3a}\includegraphics[width= 0.4\textwidth]{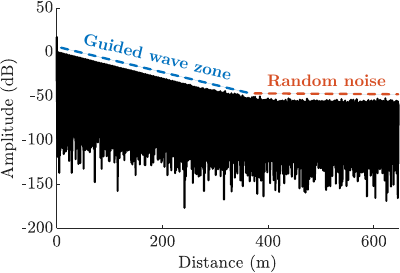}} \\
\subfloat[]{\label{Figure 3b}\includegraphics[width= 0.4\textwidth]{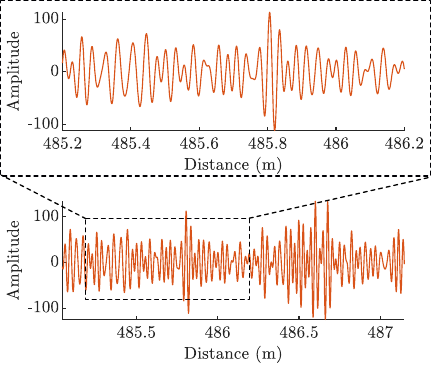}} \hspace{0.5cm}
\subfloat[]{\label{Figure 3c}\includegraphics[width= 0.4\textwidth]{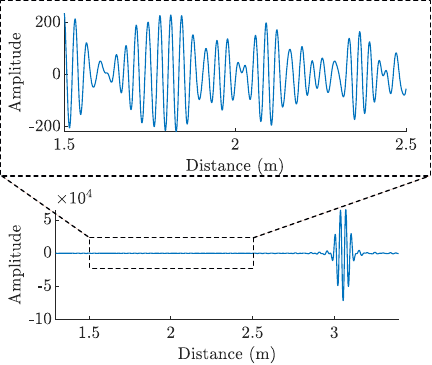}} \\ 

\caption{(a) A representative T(0,1) guided wave signal measured with an extended duration depicted on the logarithmic scale. (b) Random noise collected from the latter part of the signal where guided wave signals dissipate sufficiently to a level below that of the random noise. (c) The signal containing coherent noise and the first T(0,1) wave reflected from the pipe end.}\label{Figure 3}
\end{figure}

Figure \ref{Figure 3a} presents a representative T(0,1) guided wave signal collected on Pipe 2, depicted on the logarithmic scale and acquired with an extended duration. The signal was divided into two zones based on the amplitude on the logarithmic scale. The initial segment of the signal is dominated by guided waves, exhibiting a linear decay in logarithmic amplitude due to material damping. The latter part of the signal is dominated by random noise, where guided wave signals dissipate sufficiently to a level below that of the random noise, maintaining a constant amplitude level. The random noise signal obtained from the later part is illustrated in Fig. \ref{Figure 3b}. The accompanying zoomed-in figure in Fig. \ref{Figure 3b} displays the random noise signal within a 1 m travelling range, exhibiting randomly varying amplitudes. \par

The initial segment of the signal comprises numerous reverberations of T(0,1) reflected from pipe ends due to the short length of the pipe. Figure \ref{Figure 3c} shows the signal containing the first T(0,1) wave reflected from the pipe end at 3 m. In this segment, all components apart from the T(0,1) wavepacket reflected from the pipe end are identified as noise, comprising both coherent and random noise. A closer look in Fig. \ref{Figure 3c} reveals that the noise in this part of the signal exhibits a particular wave pattern, likely caused by the coherent noise formed by guided waves rather than purely from random noise. It should be noted that the signal in Fig. \ref{Figure 3b} was filtered by a bandpass filter, excluding the very low-frequency and very high-frequency components of random noise. Consequently, the visual appearance of the signal in Fig. \ref{Figure 3b} appears less random compared to white noise. \par

\subsection{Random noise} \label{randomNoise}

\begin{figure}
\centering 
\subfloat[]{\label{Figure 4a}\includegraphics[width= 0.4\textwidth]{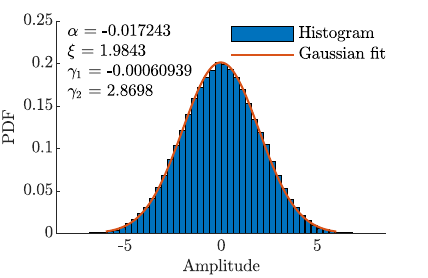}} \quad
\subfloat[]{\label{Figure 4b}\includegraphics[width= 0.4\textwidth]{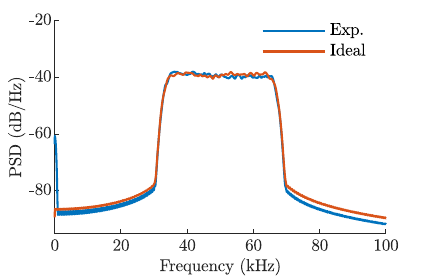}} 
\caption{Statistical analysis of the random noise. (a) PDF of the random noise alongside the fitting of an ideal Gaussian distribution, where \(\alpha\) is the mean value, \(\xi\) is the standard deviation, \(\gamma_{1}\) and \(\gamma_{2}\) are Skewness and Kurtosis, respectively. (b) PSD of the experimental random noise signal compared with that of an ideal Gaussian white noise.}
\label{Figure 4}
\end{figure}

Random noise is commonly represented using Gaussian white noise\cite{khurjekar2019accounting}, with its intensity primarily dependent on the measurement system and ambient environment. Figure \ref{Figure 4} presents an analysis of the statistical characteristics of the random noise collected in the laboratory environment. Two fundamental statistical moments, namely Skewness and Kurtosis, were employed to gauge the deviation of the collected random noise distribution from normality\cite{birnie2016analysis,martinez2015computational}. Skewness and Kurtosis are defined as
\begin{equation}
    \label{Equation 1}
    \gamma_{1} 
 = \frac{E\lbrack(x-\alpha)^3\rbrack}{E\lbrack(x-\alpha)^2\rbrack^{3/2}} ,
 \end{equation}
 and 
 \begin{equation}
    \label{Equation 2}
    \gamma_{2} 
 = \frac{E\lbrack(x-\alpha)^4\rbrack}{E\lbrack(x-\alpha)^2\rbrack^{2}} ,
 \end{equation}
where \(E\lbrack\cdot\rbrack\) denotes the expectation operator, \(x\) is the data point, and \(\alpha\) is the mean value of data. Skewness quantifies the asymmetry of the distribution around the sample mean, while Kurtosis indicates the degree of flatness near the distribution centre. For a Gaussian distribution, Skewness and Kurtosis are equal to 0 and 3, respectively.  \par

Figure. \ref{Figure 4a} presents the Probability Density Functions (PDF) of the random noise alongside the fitting of an ideal Gaussian distribution. The shape of the PDF and the Skewness and Kurtosis values imply that the collected random noise follows a Gaussian distribution.\par
To assess whether the noise power spectrum remains white, the Power Spectral Density (PSD) of the random noise signal was computed and compared with that of an ideal Gaussian white noise possessing a comparable power level, both subjected to the same bandpass filter. The results in Fig. \ref{Figure 4b} reveal that the spectrum of the recorded noise remains nearly constant within the frequency range of interest, confirming its white power spectrum.\par
\begin{figure}[hbt!]
\centering
\subfloat[]{\label{Figure 5a}\includegraphics[width= 0.4\textwidth]{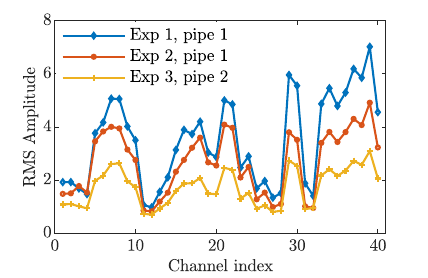}} 
\subfloat[]{\label{Figure 5b}\includegraphics[width= 0.4\textwidth]{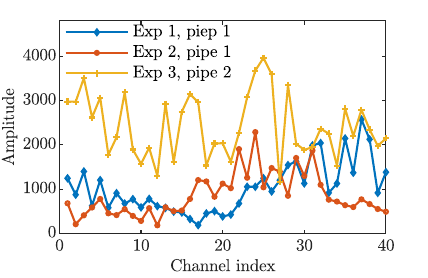}} 

  \caption{Influence of installation conditions on levels of random noise and the first T(0,1) wavepacket reflected from the pipe end. (a) RMS amplitude distribution of random noise signals from 40 transducer channels under different installation conditions. (b) The corresponding amplitude of the first T(0,1) wavepacket reflected from the pipe end from 40 transducer channels under different installation conditions.}
  \label{Figure 5}
\end{figure}

To investigate the influence of transducer installation conditions, such as the coupling between transducer elements and the pipe surface, on random noise, the transducer ring was detached and reinstalled before data collection to introduce some variability to the installation conditions. Figure \ref{Figure 5a} displays the Root Mean Square (RMS) amplitude spectrum of random noise signals from 40 transducer channels under different installation conditions. Figure \ref{Figure 5b} shows the corresponding amplitude of the first T(0,1) wavepacket reflected from the pipe end. The data were collected from Pipe 1 and Pipe 2, as specified in Table \ref{Table 1}, in different experiment sets. The RMS amplitude of random noise varies among transducer channels but maintains a consistent distribution across experiment sets. However, the amplitude of the first T(0,1) wavepacket reflected from the pipe end varies among transducer channels and changes with installation conditions in different experiment sets. This indicates that the signal-to-random noise ratio of the signal from different transducer channels depends on the transducer installation conditions. \par

Consequently, a Gaussian white noise model with limited bandwidth will be adopted to simulate random noise in the proposed numerical framework, generating realistic guided wave signals. The level of random noise at each transducer channel will be adjusted based on the first T(0,1) wavepacket reflected from the known pipe features, such as the pipe end in this study, obtained in the representative experiments.

\subsection{Coherent noise}\label{Coherent Noise}

\begin{figure}[hbt!]
\centering
\subfloat[]{\label{Figure 6a}\includegraphics{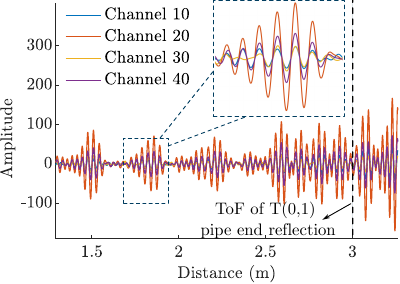}} \\
\subfloat[]{\label{Figure 6b}\includegraphics{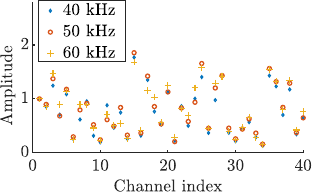}} 
\subfloat[]{\label{Figure 6c}\includegraphics{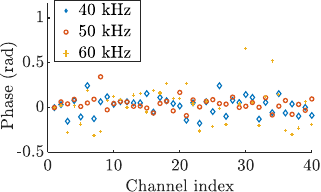}} 
\subfloat[]{\label{Figure 6d}\includegraphics{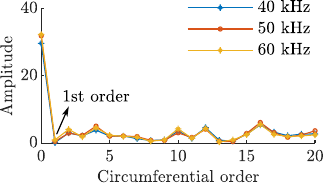}} 
\caption{(a) Pulse-echo signals from four individual transducers when each transducer was individually excited, and quantitative analysis of (b) amplitude and (c) phase variations across 40 transducer channels at three distinct frequencies. (d) Spatial Fourier transforms for analysing the excitation of guided wave modes with different circumferential orders using transducers with performance variations.}
\label{Figure 6}
\end{figure}
This section analyses the characteristics and composition of coherent noise based on the experimental data collected on Pipe 2. To mitigate the influence of random noise, 21 averages were taken in the collection of experimental signals. After averaging 21 times, the signal exhibits a consistent appearance, and the effect of random noise is effectively eliminated. \par
Coherent noise in the T(0,1) guided wave signal arises from the imperfect cancellation of circumferential waves and the excitation of unwanted flexural modes due to variations in transducer performance. For the excitation of pure T(0,1) waves, all transducers need to exhibit the same response to cancel out circumferential waves and suppress axial flexural modes. However, practical imperfections in excitation and reception, such as limited measurement system bandwidth and uneven transducer coupling conditions, inevitably introduce variations in transducer response. It is important to note that the transducer response is determined by all the electrical and electro-mechanical components present in the measurement system, including the pulser, cables, transducers, amplifiers, and transducer coupling conditions. Therefore, the discussion of transducer performance in this paper encompasses all factors that may affect the guided waves in the signal generation and reception process.\par
Figure \ref{Figure 6a} presents pulse-echo signals from four individual transducers when each transducer was individually excited. The pulse-echo signal before the first T(0,1) pipe end reflection consists primarily of circumferential waves, specifically the quasi-A0 and quasi-S0 modes. Here, the circumferential wave modes are termed quasi-S0 and quasi-A0 to distinguish them from the wave modes in a plate. The variations in the wavepackets, as illustrated in the zoomed-in figure in Fig. \ref{Figure 6a}, directly correspond to the variations in the response from an individual transducer. Quantitative analysis of amplitude and phase variations across 40 transducer channels at three distinct frequencies is depicted in Fig. \ref{Figure 6b} and Fig. \ref{Figure 6c} respectively. Amplitude and phase exhibit non-uniformity across transducer channels, while phase variations also demonstrate a dependency on frequency particularly due to the dispersion of quasi-A0 and quasi-S0 modes. These variations in transducer performance disrupt the destructive interference of circumferential waves, resulting in residual signals that manifest as coherent noise following the summation process for T(0,1) signal extraction.\par

\begin{figure}[hbt!]
\centering
\subfloat[]{\label{Figure 8a}\includegraphics{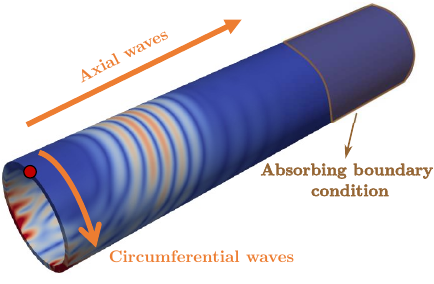}} \\
\subfloat[]{\label{Figure 8b}\includegraphics{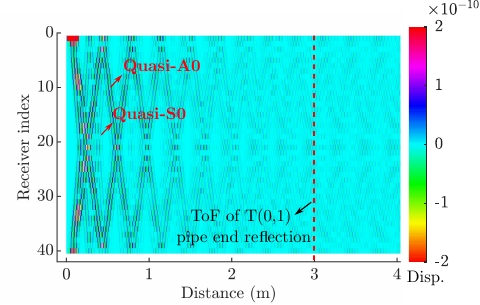}} 
\subfloat[]{\label{Figure 8c}\includegraphics{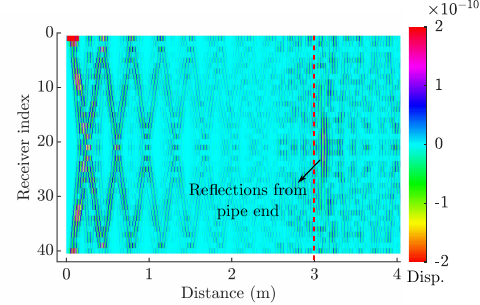}} \\
\subfloat[]{\label{Figure 8d}\includegraphics{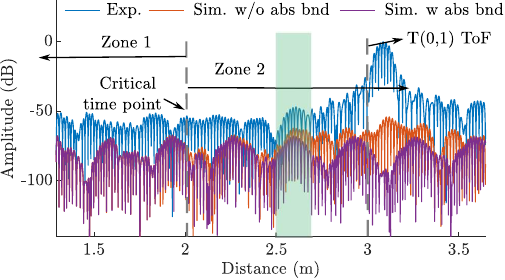}} 
\subfloat[]{\label{Figure 8e}\includegraphics{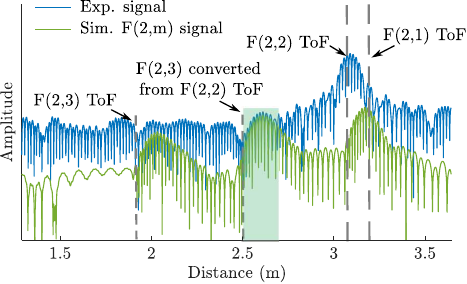}} 
\caption{(a) Circumferential waves and axial waves propagating in a pipe with a point source excitation. (b) Simulated amplitude map of signals from all receivers when transducer No. 1 was excited, with an absorbing boundary applied in the simulation. (c) Simulated amplitude map of signals from all receivers when transducer No. 1 was excited, without absorbing boundary. (d) Comparison of the simulated time-trace signals from transducer No. 20 under two different boundary conditions with the experimental T(0,1) signal with coherent noise. (e) Comparison of the extracted F(2,m) modes with the experimental T(0,1) signal with coherent noise.}
\label{Figure 8}
\end{figure}

A spatial Fourier transform along the circumferential direction of the pipe was performed to analyse the excitation of the guided wave modes using transducers with the performance variations shown in Fig. \ref{Figure 6b}. The amplitude spectrum of the spatial Fourier transform reflects the possibility of exciting an axial guided wave mode with different circumferential orders, denoted as F(n,m), where n represents the circumferential order, and m represents the counter index of a flexural wave. Figure \ref{Figure 6d} displays the amplitude spectrum at three frequencies. The low amplitude of order 1 suggests that the F(1,m) mode is not excited with the transducer performance imbalance depicted in Fig. \ref{Figure 6b}.

The contribution of circumferential waves and axial flexural waves to coherent noise is analysed here. Figure \ref{Figure 8a} shows a simulation of the wavefield generated by a single point source, consisting of waves propagating in the circumferential direction and axial direction. The contribution of circumferential waves and axial flexural waves can be identified by comparing the amplitude map of signals from all receivers in simulations with different boundary conditions. In the simulation, 40 transducers were located at one end of the pipe, serving as transmitters as well as receivers. Figure \ref{Figure 8b} illustrates the amplitude map of signals from all 40 receivers excited by transducer No. 1 with an absorbing boundary applied in the simulation, as shown in Fig. \ref{Figure 8a}. The excitation was simulated by a point source with a shear force applied to the pipe surface in the circumferential direction. In this case, the signals received by receivers consist only of circumferential waves, observed as the quasi-S0 mode and the slower quasi-A0 mode with higher amplitude propagating circumferentially. In contrast, Fig. \ref{Figure 8c} presents the amplitude map of signals from all 40 receivers excited by transducer No. 1 without an absorbing boundary. Here, axial waves reflected from the pipe end can be sensed by receivers, contributing to the flexural modes propagating axially. \par
To reveal the contribution of circumferential waves and axial flexural waves to coherent noise, the simulated time-trace signals from transducer No. 20 under two different boundary conditions are compared with the experimental T(0,1) signal in Fig. \ref{Figure 8d} on a logarithmic scale. Figure \ref{Figure 8d} shows two distinct zones separated by a critical time point, where only circumferential waves are present in Zone 1, and both circumferential and axial waves are present in Zone 2. The amplitude of circumferential waves diminishes inversely as the square root of the distance from the point source location due to the geometrical spreading in what is effectively a two-dimensional space. Therefore, the contribution of circumferential waves to coherent noise decreases with the wave propagation distance due to the energy spread.\par

In Zone 2, a coherent wavepacket is visible in the coherent noise (marked by a green area), which is contributed to by both circumferential waves and axial flexural waves. The mode of the contributing axial flexural wave can be identified by comparing the time of flight (ToF) of each mode with the experimental signal. In the experiments, shear force was applied to the pipe circumference, mainly resulting in the excitation of shear displacement-dominated wave modes, such as F(1,2), F(2,2), and F(3,2), as shown in Fig. \ref{Figure 7}. As mentioned before, there is no F(1,m) mode present when transducer performance follows the variations shown in Fig. \ref{Figure 6b}. Additionally, the energy of higher flexural modes like F(3,m) is relatively small compared to the lower-order modes\cite{hayashi2005mode}. Thus, only F(2,m) was considered and compared with the experimental T(0,1) signal to identify the wave modes contributing to the coherent wavepacket marked in Zone 2.\par

For the calculation of the ToFs of each wave mode, considering mode conversion on the pipe end, three-dimensional finite element simulations were employed based on Pipe 2. For the excitation and extraction of a specific flexural mode F(n,m), a phase shift corresponding to the circumferential order n was applied to each node for both excitation and reception before summing the signals\cite{Mike10.1115/1.2789107}. Figure \ref{Figure 8e} illustrates the comparison of extracted F(2,m) modes with the experimental T(0,1) signal on a logarithmic scale. A high similarity in the ToF between the F(2,3) wavepacket, converted from the F(2,2) mode, and the marked wavepacket was observed. It means this coherent wavepacket is likely produced by circumferential waves and the F(2,3) mode converted from F(2,2) on the pipe end. It is worth noting that the amplitude of the F(2, m) mode in Fig. \ref{Figure 8e} is depicted on an arbitrary scale, and the quantitative amplitude of a flexural mode depends on specific installation conditions. The qualitative analysis of this falls beyond the scope of this study.\par

\section{Methodology}\label{Methodology}
\subsection{Generic scheme}\label{Generic scheme}

\begin{figure}[!t]
\centerline{\includegraphics{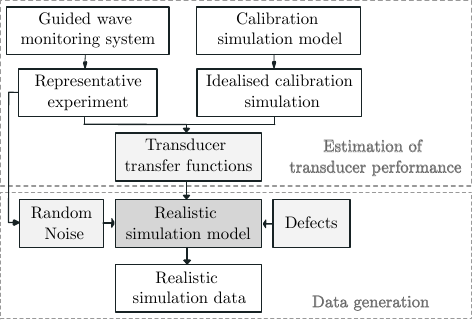}}
\caption{Schematic of the proposed framework for generating realistic guided wave signals through FE simulation.}
\label{Figure 9}
\end{figure}

In this section, we introduce a numerical framework developed for generating synthetic realistic guided wave signals, which relies on a single representative experiment. The schematic diagram of the proposed methodology is illustrated in Fig. \ref{Figure 9}. To faithfully represent experimental signals, our framework incorporates both coherent noise and random noise into the simulation.\par
In the proposed framework, simulations are conducted in two key phases: the idealised calibration simulation and the realistic simulation. In the idealised calibration simulation, all transducers are assumed to have uniform performance. A CSM dataset is obtained by applying the identical excitation signals to all transmitter nodes and the T(0,1) signal is obtained by summing the CMS signals from the reception side. In the realistic simulation model, both coherent noise and random noise are incorporated to generate realistic guided wave signals. For coherent noise consideration, transducer transfer functions are estimated and applied in the simulation on both the excitation and reception sides. Subsequently, random noise is added to the signals from each transducer channel. A realistic T(0,1) signal can be obtained by summing the signals from all transducer channels.\par

The incorporation of random noise involves a statistical analysis of the noise collected in the representative experiments, which is specific to the ambient environments and measurement system employed. In this study, random noise can be effectively modelled using Gaussian white noise with a limited bandwidth, as described in \ref{randomNoise}. The RMS level of the random noise across transducer channels follows the distribution described in \ref{Figure 5a}.\par
To model coherent noise, transducer transfer functions are estimated based on a comparison between the representative experimental signal and the idealised calibration simulation signal in the frequency domain. The transducer transfer function obtained in this manner has shown good agreement with the transfer function obtained by characterising all electrical and electro-mechanical components within the measurement system in the literature\cite{sanchez2006characterization,lopez2005ultrasonic}. The implementation of these transfer functions enables the capture of the transducer performance across the entire frequency range of interest, which is necessary for the representation of coherent noise, as discussed in Section \ref{Coherent Noise}. This section will illustrate the methodology for estimating the transducer transfer functions in detail.\par
Without considering the effect of transducer performance, the signal emitted from transducer \(i\) and received by transducer \(j\) is denoted as \(S_{ij}\), and represents the physical wave propagation in both the circumferential and axial directions, as illustrated in Fig. \ref{Figure 8a}. The T(0,1) signal, \(S_{\mathrm{T01}}\), is obtained by summing the signals from all transducer pairs, as given by    
\begin{equation}
    \label{Equation 3 - ideal sum up}
    S_{\mathrm{T01}} = \sum_{i=1}^{N}\sum_{j=1}^{N} S_{ij} ,
\end{equation}
where \(N\) represents the number of transducer elements. In an ideal measurement scenario, \(N\) transducers equally spaced around the pipe's circumferential direction are excited with identical signals. The circumferential waves are suppressed through destructive interference, resulting in the formation of a clear T(0,1) plane wave propagating axially. Simultaneously, the enforced identical displacement around the pipe circumference induced by \(N\) transducers suppresses all flexural modes with a circumferential order lower than \(N\), leaving only the T(0,1) wave mode propagating along the pipe axis within the frequency range of interest. In practical situations, both coherent noise and random noise exist in the T(0,1) signal, which can be described as
\begin{equation}
    \label{Equation 4 - practical sum up}
    S'_{\mathrm{T01}} = \sum_{i=1}^{N}\sum_{j=1}^{N} S'_{ij} ,
\end{equation}
where \(S'_{ij}\) represents the signal \(S_{ij}\) contaminated by random noise and affected by transducer performance. 
The transducer performance can be represented as transducer transfer functions, \(T_{ti}\) and \(T_{rj}\), on the excitation and reception sides, respectively, within the frequency range of interest. Hence, \(S'_{ij}\) can be expressed as 
\begin{equation}
    \label{Equation 5 -single practical signal}
    S'_{ij} = T_{ti}\cdot T_{rj}\cdot S_{ij}+S'_{\mathrm{N}ij} ,
\end{equation}
where \(S'_{\mathrm{N}ij}\) represents the random noise signal in the signal \(S'_{ij}\) and \(\cdot\) donates multiplication in the frequency domain, corresponding to convolution in the time domain. Equation (\ref{Equation 4 - practical sum up}) is transformed to  
\begin{equation}
    \label{Equation 6 - practical sum up ij}
    S'_{\mathrm{T01}} = \sum_{i=1}^{N}\sum_{j=1}^{N} (T_{ti}\cdot T_{rj}\cdot S_{ij}+S'_{\mathrm{N}ij}) .
\end{equation}
\par
In Eq. (\ref{Equation 6 - practical sum up ij}), \(S_{ij}\), representing the physical wave propagation, can be accurately captured in an idealised calibration simulation. The transducer transfer functions \(T_{ti}\) and \(T_{rj}\) can be estimated based on the T(0,1) wavepacket reflected from the known features, such as welds, collected in the representative experiment. \par

When data is captured with the CSM arrangement, Eq. (\ref{Equation 6 - practical sum up ij}) can be alternatively expressed as
\begin{equation}
    \label{Equation 10 - practical sum up in - CSM}
    S'_{\mathrm{T01}} = \sum_{j=1}^{N} (T_{rj}\cdot \sum_{i=1}^{N} (T_{ti}\cdot S_{ij})+S'_{\mathrm{N}j}) ,
\end{equation}
where \(S'_{\mathrm{N}j}\) represents the random noise signal.
The signal available from transducer \(j\) in an experimental measurement is denoted as 
\begin{equation}
    \label{Equation 11 - practical sumup-i-CSM}
    S'_{j} = T_{rj}\cdot \sum_{i=1}^{N} (T_{ti}\cdot S_{ij})+S'_{\mathrm{N}j} .
\end{equation}
The plane wave formed by the common source excitation is assumed to be minimally affected by the performance imbalance of transducers on the excitation side due to wave superposition. The plane wave consists of both T(0,1) and other unwanted flexural modes resulting from the imbalance of transducer performance. For cases where the functional transducer ring is properly installed under the guided wave monitoring guidelines, the amplitude of the unwanted flexural modes mixed in the T(0,1) wavepacket used for the estimation of a transducer transfer function is negligible compared to the amplitude of the T(0,1) wavepacket, due to the limited transducer performance variations and the mode conversion that occurs at pipe feature locations. It allows for the assumption that variations on the excitation side can be omitted, resulting in
\begin{equation}
    \label{Equation 12 - practical sumup-i-CSM_approx}
    S'_{j} \approx T_{rj}\cdot \sum_{i=1}^{N} S_{ij} +S'_{\mathrm{N}j} .
\end{equation}
Hence, the one-side transfer function of transducer \(j\) can be estimated as: 
\begin{equation}
    \label{Equation 13 - TransFun-i-CSM_approx}
    T_{rj} \approx \frac{S'_{j}}{S_{j}} ,
\end{equation} 
when the random noise \(S'_{\mathrm{N}j}\) is negligible, where \(S_{j} = \sum_{i=1}^{N} S_{ij}\) is the receiving signal from transducer \(j\) in the calibration simulation. \par

Therefore, the transducer transfer functions can be estimated by Eq. (\ref{Equation 13 - TransFun-i-CSM_approx}) based on the comparison of the idealised calibration simulation signal and the representative experimental signal. It is important to note that the direct division in these equations is unstable due to the inherent presence of random noise. To address this, a Wiener filter can be implemented to stabilize the estimation of transducer transfer functions\cite{wiener1964wiener}. Equation (\ref{Equation 13 - TransFun-i-CSM_approx}) can be reformulated as:  
\begin{equation}
    \label{Equation 13 - Winner filter of equation 7}
    T_{rj} \approx \frac{S^*_{j}\cdot S'_{j}}{|S_{j}|^2+\lambda} ,
\end{equation}
where \(*\) denotes the complex conjugate and \(\lambda\) represents the stabilization factor, which is determined based on the noise level of the signal \(S'_{j}\) and can be estimated by
\begin{equation}
    \label{Equation 14 -Factor in winner filter}
    \lambda = K|S_{j}|^2_{max} .
\end{equation}
Here, the parameter \(K\) is determined empirically according to the noise level of the signal \(S'_{j}\). A smaller \(K\) value may result in inadequate noise suppression, whereas a larger value can lead to a higher error in the estimation of transducer transfer functions\cite{cicero2009potential}. In this study, \(K\) has been set as 0.0005 through a trial-and-error procedure. \par

\subsection{Simulation setup}
All simulations in this study were performed using Pogo, an FE solver developed at Imperial College London\cite{huthwaite2014accelerated}. Three-dimensional FE models were constructed based on the specific pipes employed in the experiments, as detailed in Table \ref{Table 1}. In the simulation, the transducer ring was positioned at the end of the pipe, with 40 equally spaced nodes representing 40 transducer elements for guided wave excitation and reception.\par 
Aligned with the experimental setup, a 5-cycle Hann-windowed toneburst with a central frequency of 50 kHz was used as the excitation signal in the simulation, characterised by a T(0,1) wavelength of approximately 60 mm. To ensure a precise simulation of wave propagation, a structured mesh with an element size of 1 mm, comprising roughly 60 elements per wavelength, was implemented, with a corresponding time step of approximately \(1\times 10^{-7}\) s. The total simulation duration was 10 ms, capturing the first six T(0,1) reflections from the pipe end. The chosen element type was a general-purpose linear brick element with reduced integration (C3D8R). For a typical simulation of Pipe 1, the number of nodes was 16,691,840, requiring 6,858.92 MB of memory and approximately 7 minutes for a 10 ms wave propagation simulation on three NVIDIA GeForce RTX 2080 Ti GPUs.\par

\subsection{Simulation validation}\label{simualtion_validation}

To validate the methodology proposed in this paper for the estimation of transducer transfer functions, we generated a set of reference simulation signals with coherent noise based on Pipe 1, using the same excitation configuration as in the experiment. These signals were then stored for comparison as the ground truth, serving as the experimental data within the framework illustrated in Fig. \ref{Figure 9}. The process described in Section \ref{Generic scheme} was applied to estimate transducer transfer functions. Notably, random noise was excluded in this section, with specific emphasis placed on coherent noise.\par

\begin{figure}
\centering 
\subfloat[]{\label{Figure 10a}\includegraphics{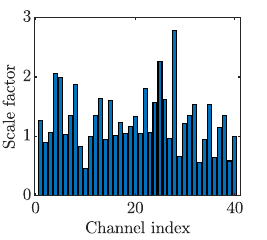}} 
\subfloat[]{\label{Figure 10b}\includegraphics{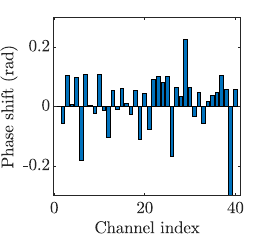}} 
\subfloat[]{\label{Figure 10c}\includegraphics{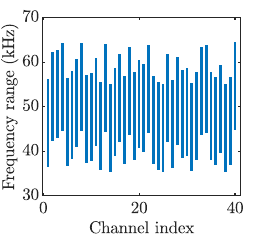}} \\
\subfloat[]{\label{Figure 10d}\includegraphics{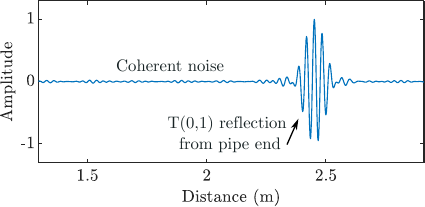}} 
\caption{Distribution of (a) phase shifts, (b) amplitude variations, and (c) bandwidths across transducer channels applied in the reference simulation. (d) T(0,1) signal with coherent noise generated in the reference simulation.}
\label{Figure 10}
\end{figure}

For the generation of reference simulation signals, different phase shifts, amplitude variations, and varying passbands were applied to both the excitation and reception sides, as depicted in Fig. \ref{Figure 10a} to Fig. \ref{Figure 10c}. The resulting reference T(0,1) signal is presented in Fig. \ref{Figure 10d} with coherent noise present.\par

\begin{figure}[!t]   
\centerline{\includegraphics{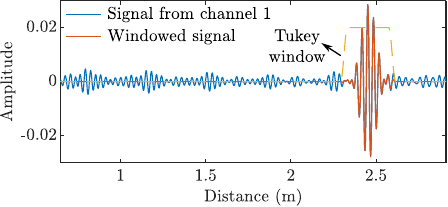}}
\caption{Reference signal \(S’_{1}\) (shown in orange) acquired from transducer No.1 used for estimating transducer transfer functions, which were windowed from the original signals (shown in blue) with a Tukey window indicated by the yellow dotted line. }
\label{Figure 11}
\end{figure}

\begin{figure}
\centering 

\subfloat[]{\label{Figure 12a}\includegraphics{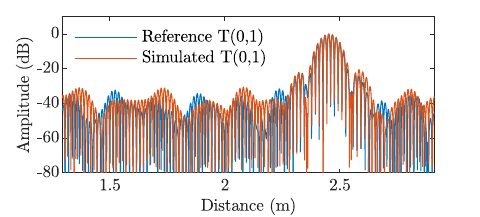}} 
\subfloat[]{\label{Figure 12b}\includegraphics{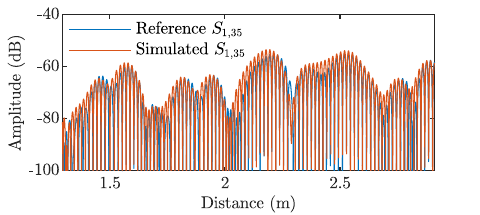}} \\
\subfloat[]{\label{Figure 12c}\includegraphics{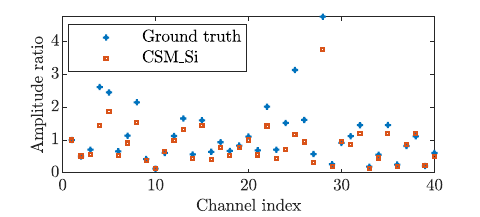}} 
\caption{(a) Comparison of the simulation-generated T(0,1) signal with the T(0,1) reference signal on the logarithmic scale. (b) Comparisons between the simulation-generated pitch-catch signal emitted from transducer No. 1 and received by transducer No. 35, with the corresponding reference signal \(S'_{1,35}\). (c) Comparisons between the estimated amplitude variations with the ground truth of amplitude variations applied in the reference simulation.}
\label{Figure 12}
\end{figure}

The one-side transfer function of transducer \(i\) was estimated using Eq. (\ref{Equation 13 - Winner filter of equation 7}) based on the receiving signal \(S'_{i}\) from transducer \(i\) with the CSM data capture arrangement. Figure \ref{Figure 11} illustrates the reference simulation signal acquired from transducer No. 1, \(S'_{1}\), when all transducers were excited simultaneously. The T(0,1) wavepacket reflected from the pipe end was windowed out using a Tukey window for the estimation of the transducer transfer function. \par

The simulation-generated T(0,1) signal with coherent noise is compared with the reference T(0,1) signal in Fig. \ref{Figure 12a}. The simulation-generated T(0,1) signal shows a similar coherent noise level as that in the reference T(0,1) signal but with some discrepancies in the waveform. Furthermore, Fig. \ref{Figure 12b} shows the simulation-generated pitch-catch signal emitted from transducer No. 1 and received by transducer No. 35, which is overestimated compared to the corresponding reference signal \(S'_{1,35}\). The discrepancy in the simulation-generated signal and the reference signal is due to the approximation in the estimation of transducer transfer functions made in Eq. (\ref{Equation 12 - practical sumup-i-CSM_approx}).\par
 
Figure \ref{Figure 12c} compares the estimated amplitude variations with the ground truth of amplitude variations applied in the reference simulation. The estimation generally shows good agreement with the ground truth but exhibits a larger deviation when the variations in transducer response become too large, thereby violating the assumption in Eq. (\ref{Equation 12 - practical sumup-i-CSM_approx}).

\subsection{Experimental validation}

In this section, the proposed methodology is evaluated based on experimental signals collected from Pipe 2. As mentioned in Section \ref{Coherent Noise}, the experimental signals were obtained with 21 averages to suppress random noise, focusing on the effects of coherent noise.

\subsubsection{Individual T(0,1) signal comparison}
\begin{figure}[hbt!]
\centering
\subfloat[]{\label{Figure 13a}\includegraphics{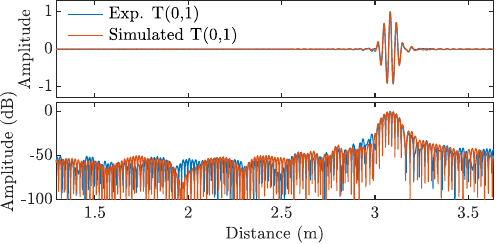}} 
\subfloat[]{\label{Figure 13b}\includegraphics{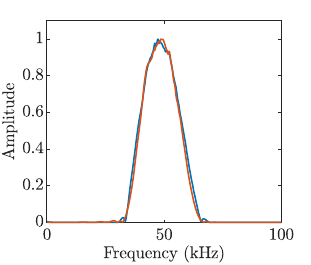}} \\

\caption{Comparison between the simulation-generated T(0,1) signals and the corresponding experimental T(0,1) signals in (a) the time domain and (b) the frequency domain.}
\label{Figure 13}
\end{figure}

Figure \ref{Figure 13a} displays the simulation-generated T(0,1) signal based on the estimated transducer transfer functions. The comparison with the experimental T(0,1) signal reveals a good agreement in both the coherent noise level and the waveform of the T(0,1) wavepacket reflected from the pipe end. The frequency spectrum comparison in Fig. \ref{Figure 13b} highlights the similarity between the simulation-generated signal and the experimental T(0,1) signal, displaying an overall smooth shape with minor distortions attributed to the presence of coherent noise.\par

\subsubsection{Error analysis}

Similarly to the comparison in the simulation validation shown in Fig. \ref{Figure 12a}, the comparison between the simulation-generated T(0,1) signal and the experimental signals in Fig. \ref{Figure 13a} presents a discrepancy in the waveform. One possible reason for this discrepancy is the error in estimating transducer transfer functions. Assumptions were made about random noise and the variations in transducer responses in Eq. (\ref{Equation 13 - Winner filter of equation 7}) for the estimation of transducer transfer functions. These assumptions may be violated for certain transducer channels where random noise or transducer response variations are more pronounced. In the experiments, additional error sources may be present due to imperfections in the real world. For instance, variations in pipe wall thickness\cite{KARA2010998} and material anisotropy resulting from the manufacturing process\cite{pyshmintsev2016evolution} can contribute to these errors. The manufacturing of seamless pipes typically involves processes such as extrusion and hot finishing, which inevitably introduce changes in the material microstructure, leading to anisotropy, particularly in the axial and circumferential directions\cite{malachowski2014investigation}. These imperfections introduce extra errors in the idealized calibration simulation signal \(S_{j}\) presented in Eq. (\ref{Equation 13 - Winner filter of equation 7}).\par

  \begin{figure}[hbt!]
\centering   
\subfloat[]{\label{Figure 15a}\includegraphics{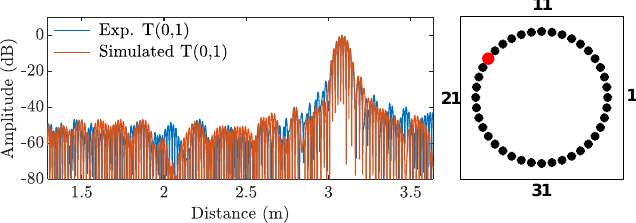}}  \\
\subfloat[]{\label{Figure 15b}\includegraphics{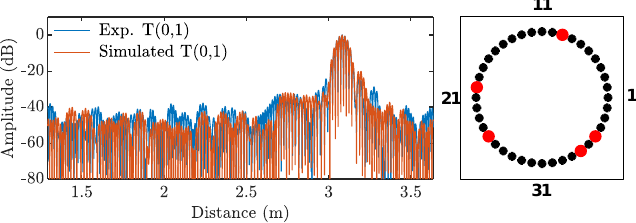}} \\
\subfloat[]{\label{Figure 15c}\includegraphics{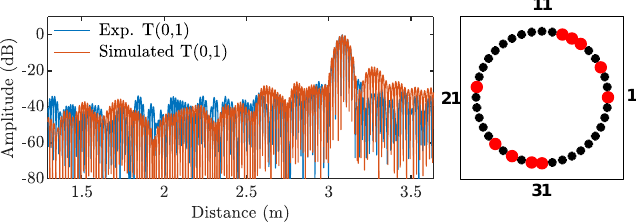}} 
\caption{Comparison between the simulation-generated T(0,1) signals (shown in orange) with the experimental T(0,1) signals (shown in blue) with different coherent noise levels induced by the disabled transducer channels, which are highlighted in red in the corresponding figures shown on the right. (a) One transducer channel disabled. (b) Five transducer channels disabled. (c) Ten transducer channels disabled.}
\label{Figure 15}
\end{figure}  

\subsubsection{Statistical T(0,1) signal comparison}

In this section, we conduct a statistical analysis to quantitatively evaluate the effectiveness of the proposed framework in accurately representing coherent noise. Experimental data with various transducer performance variations, hence exhibiting different coherent noise levels, were collected based on a full-matrix dataset obtained by exciting individual transmitters sequentially. Specifically, a varying number of transducer channels were randomly selected and disabled, and the corresponding signals in the full-matrix dataset related to these channels were suppressed. Generally, an increase in the number of disabled transducer channels results in a higher coherent noise level in the experimental signal. Subsequently, the modified full-matrix dataset was summed on the excitation side to acquire the CSM data, which was then used to estimate transducer transfer functions according to the method detailed in Section \ref{Generic scheme}. \par
\begin{figure}[ht]
\centerline{\includegraphics{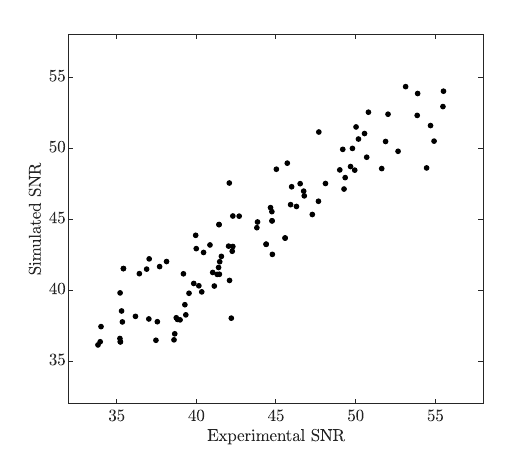}}
\caption{Scatter plot of the SNR values of the simulation-generated T(0,1) signals against those of the experimental T(0,1) signals, comprising 100 data samples.}
\label{Figure 16}
\end{figure}

Figure \ref{Figure 15} compares the simulation-generated T(0,1) signals with the experimental T(0,1) signals, showcasing diverse coherent noise levels. The disabled transducer channels are highlighted in red in the corresponding figures on the right side. Overall, a good agreement is observed in the coherent noise level. For a quantitative evaluation, the coherent noise level of a T(0,1) signal was calculated using signal-to-noise ratio (SNR), defined as:
\begin{equation}
    \label{Equation 16}
    \text{SNR} = 20log_{10}(\frac{RMS_{\mathrm{Coherent Noise}}}{AMP_{\mathrm{T(0,1)}}}) ,
\end{equation}
where \(RMS_{\mathrm{Coherent Noise}}\) represents the RMS value of coherent noise, and \( AMP_{\mathrm{T(0,1)}}\) denotes the amplitude of the T(0,1) wavepacket reflected from the pipe end. \par

Figure \ref{Figure 16} displays a scatter plot of the SNR values of the simulation-generated T(0,1) signals against those of the experimental T(0,1) signals, comprising 100 data samples. The strong linearity observed in the SNR values between these two datasets suggests that the proposed framework performs well in quantitatively representing coherent noise in T(0,1) guided wave signals. Furthermore, the large set of realistic simulation data generated by the proposed framework can be used for various applications that require such large datasets, such as the training of machine learning models for damage identification from complex guided wave signals and POD analysis.\par
In the next section, an application in the determination of the inspection range in pipe monitoring will be illustrated based on the POD analysis using the simulation-generated realistic data.

\section{Reliability estimation}\label{Reliability estimation}

In current industrial practice, the inspection range of a guided wave system for monitoring is determined based on the system sensitivity and noise levels, as illustrated in Fig. \ref{Figure 1}. Specifically, the end of the detection/inspection is set as the point where the Call DAC drops below the Detection Threshold. As discussed in Section \ref{Real guided wave signals and noise}, the DAC curve can be quantitatively established using the Absolute Calibration method based on known weld size parameters\cite{vogt2022multiple}. However, the setting of the Detection Threshold relies on the subjective judgment of inspectors to determine the noise level for a given inspection.

In this section, we present a quantitative estimation method for determining the inspection range of a guided wave monitoring system without human intervention, using guided wave signals generated in the proposed realistic simulation model.\par

\subsection{POD analysis procedure}
For the quantitative evaluation of the detection performance of a guided wave monitoring system at a specific location, a POD analysis was conducted using the simulation-generated guided wave signals. The methodology for the POD analysis employed in this study adheres to the guidelines outlined in MIL-HDBK-1823A 2009 (Nondestructive Evaluation System Reliability Assessment)\cite{hdbk2009nondestructive}, which is a universally accepted reliability assessment guideline for NDT. It provides comprehensive instructions on performing POD analysis based on signal response and hit-miss data. In this study, the POD analysis was carried out based on the signal response data. \par
\begin{figure}[ht] 
\centerline{\includegraphics{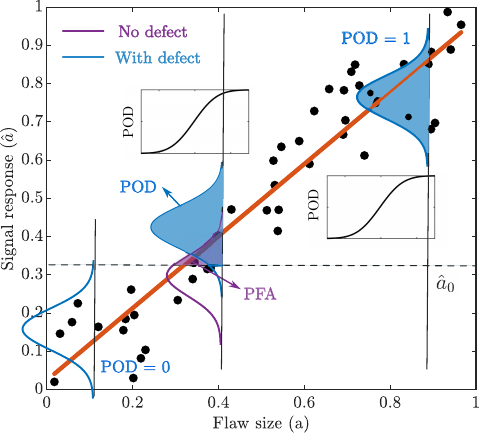}}
\caption{Illustration of the definition and calculation of a POD curve (adapted from Figure 9 on Page 28 in\cite{gandossi2010probability}).}
\label{Figure 17}
\end{figure}

POD curves illustrate how the detection probability varies with the defect parameter, such as flaw size, effectively quantifying the system’s capability to detect defects when they are present. The POD value, represented as the blue shaded area in Fig. \ref{Figure 17}, is defined as
\begin{equation}
    \label{Equation 17-Definition of POD}
    \text{POD} = P(\hat{a}>\hat{a}_{0}|\textrm{defect}) ,
\end{equation}
where \(\hat{a}\) represents the signal response, and \(\hat{a}_{0}\) denotes the Detection Threshold. 
Another metric used to assess system performance in defect detection is the Probability of False Alarm (PFA), which indicates the system’s likelihood of falsely identifying a defect when there is no defect present. The purple shaded area in Fig. \ref{Figure 17} illustrates this, and it is defined as 
\begin{equation}
    \label{Equation 18-Definition of PFA}
    \text{PFA} = P(\hat{a}>\hat{a}_{0}|\textrm{no defect}) .
\end{equation}
\par

Traditionally, the POD curve based on the signal response can be calculated using linear regression. This method assumes that the relationship between the signal response and the flaw size is linear and that any data scatter follows the same probability distribution. The parametric model can be expressed as:
\begin{equation}
    \label{Equation 19-Linear relation}
    \hat{a}_{i} = \beta_{0}+\beta_{1}a_{i}+\varepsilon_{i} .
\end{equation}
Here, \(a_{i}\) and \(\hat{a}_{i}\) represent the flaw size and the corresponding signal response measured from the experimental or simulation data. The model parameters \(\beta_{0}\), \(\beta_{1}\) and \(\varepsilon_{i}\) can be estimated based on the scattered data \((a_{i},\hat{a}_{i})\). \(\beta_{0}\) and \(\beta_{1}\) represent the intercept and slope, respectively, while \(\varepsilon_{i}\) is the random component following a normal distribution with a mean of zero and a standard deviation of \(\tau\), i.e., \(\varepsilon \sim N(0,\tau)\). The parameter \(\varepsilon_{i}\) accounts for the uncertainty of the data collected in the experiments or simulation. Estimation of model parameters can be performed through linear regression, such as ordinary least-squares (OLS) regression.\par
The shape of the POD curve can be described by the cumulative density function of a normal distribution, given by
\begin{equation}
    \label{Equation 20-normal distribution}
    p(x|\mu,\sigma) =\frac{1}{\sigma\sqrt{2\pi}} e^{-\frac{1}{2}(\frac{x-\mu}{\sigma})} .
\end{equation}
The parameters in Eq. (\ref{Equation 20-normal distribution}) are related to the parameters in Eq. (\ref{Equation 19-Linear relation}) as follows:
\begin{equation}
    \label{Equation 21-location parameter}
    \mu = \frac{a_{0}-\beta_{0}}{\beta_{1}}=a_{50} ,
\end{equation}
and 
\begin{equation}
    \label{Equation 22-shape parameter}
    \sigma = \frac{\tau}{\beta_{1}} ,
\end{equation}
where \(a_{50}\) corresponds to the flaw size where the POD is 50\%. Hence, the expression of POD is

\begin{equation}
    \label{Equation 23-POD}
    \begin{split}
    \text{POD}(a_{i}) &= P(\hat{a}_{i}>\hat{a}_{0})\\ &=1-\Phi_{norm}(\frac{\hat{a}_{0}-(\beta_{0}+\beta_{1}a_{i})}{\tau})
    \\ &= \Phi_{norm}(\frac{a_{i}-\mu}{\sigma}) ,
    \end{split}
\end{equation}
where \(\Phi_{norm}\) represents the standard normal cumulative density function. \par
With the estimated parameters based on the collected data, a confidence bound reflecting the error due to the limited sample size can be calculated using the Wald method based on the standard deviation in the estimated parameters\cite{gandossi2010probability}.

\subsection{Generation of datasets}

\newcolumntype{P}[1]{>{\centering\arraybackslash}p{#1}}
\newcolumntype{M}[1]{>{\centering\arraybackslash}m{#1}}
\begin{table*}[!h]
\renewcommand{\arraystretch}{2}
\caption{Geometric and material information of the pipe modelled for POD analysis.}

\label{Table 2}
\begin{center}
\begin{tabular}{M{0.9cm}|M{1.2cm}|M{1.0cm}|M{1.2cm}|M{1.4cm}|M{1.2cm}|M{1.2cm}|M{1.2cm}|M{1.2cm}|M{1.2cm}|M{2cm}}

   \hline\hline
\rule{0pt}{15pt}\textbf{} & \multicolumn{5}{ c |}{\textbf{Geometric}} & \multirow{3.5}{*}{\textbf{Material}}   & \multicolumn{4}{ c }{\textbf{Material Properties}}\\ 
\cline{2-6}\cline{8-11}
\rule{0pt}{18pt} &\textbf{Nominal Size (inches)}&\textbf{Schedule}  & \textbf{Outside Diameter (mm)} & \textbf{Wall Thickness (mm)} & \textbf{Length (mm)} & & \textbf{Density (kg/m$^3$)} & \textbf{Young's Modulus (GPa)} & \textbf{Poisson's Ratio}& \textbf{Mass Proportional Damping Coefficient (1/second)} \\
\hline
\rule{0pt}{15pt} Pipe 3& 5 & 40 & 141.3 & 6.55 & 10000&P265GH& 7850&210.86&0.285&1500\\ 
   \hline\hline
\end{tabular}
\end{center}
\end{table*}

\begin{figure}
\centering 
\subfloat[]{\label{Figure 18a}\includegraphics{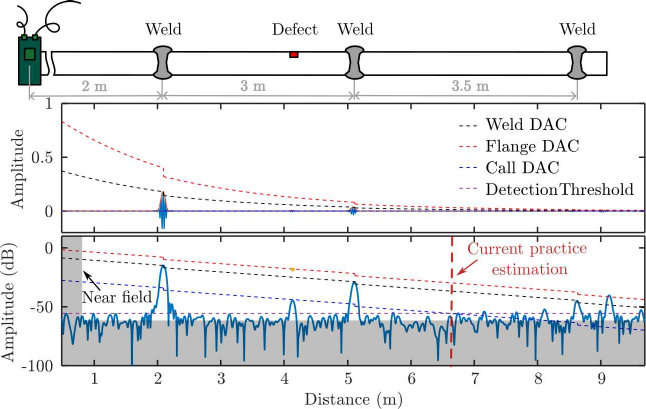}} \\
\subfloat[]{\label{Figure 18b}\includegraphics{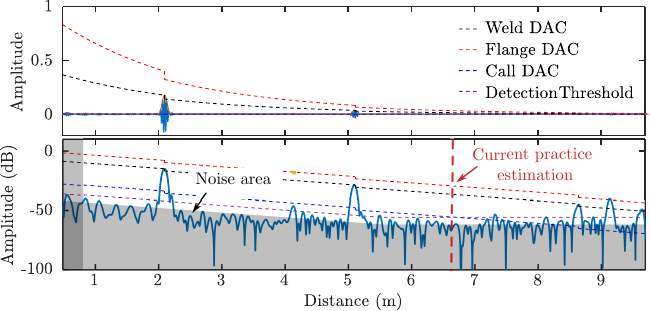}} 
\caption{Example T(0,1) signals generated in the simulation depicted on both linear (upper figure) and logarithmic (lower figure) scales. (a) T(0,1) signal with only random noise, (b) T(0,1) signal with both coherent noise and random noise.}
\label{Figure 18}
\end{figure}

\begin{figure}
\centering 
\subfloat[]{\label{Figure 19a}\includegraphics{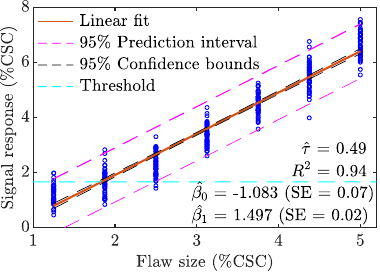}} \hspace{0.5cm}
\subfloat[]{\label{Figure 19b}\includegraphics{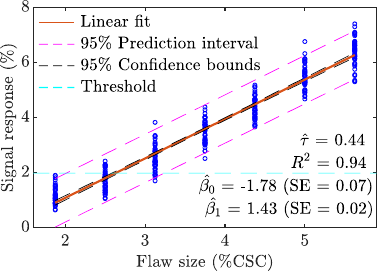}} \\
\subfloat[]{\label{Figure 19c}\includegraphics{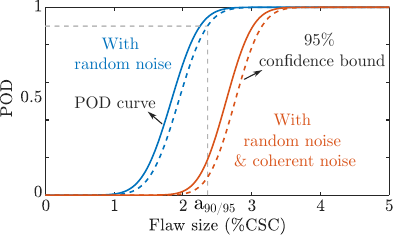}} 
\caption{Linear regression relationship between the signal response and flaw size in terms of CSC and the corresponding POD curves when the defect is positioned at a distance of 3.54 m from the transducer ring. (a) Linear regression of the data with only random noise, (b) linear regression of the data with both coherent noise and random noise, (c) comparison of the POD curves calculated based on two distinct datasets.}
\label{Figure 19}
\end{figure}

To demonstrate the quantitative estimation method for determining the inspection range of a guided wave monitoring system in a practical scenario, a series of guided wave signals were generated using the proposed numerical framework. The simulation involved a 10-m 5-inch Schedule 40 steel pipe with 3 welds, designated as Pipe 3 in Table \ref{Table 2}. To simulate the decay in amplitude caused by wave propagation, a high attenuation coefficient of 1500/second was implemented to represent a high-damping environment, such as when the pipe is buried underground. In our model, the welds were characterized by a width of 20 mm and a cap height of 3 mm, resulting in approximately 49\% CSC and a reflection coefficient of about 41\% in the absence of damping. It should be emphasized that the weld sizes in terms of CSC were exaggerated for illustrative purposes, exceeding practical norms. The positions of the welds are depicted in Fig. \ref{Figure 18}, and defects were introduced as through-thickness defects with varying circumferential extent, reaching up to 6\% CSC. An absorbing boundary condition was applied on the left side of the transducer ring, meaning only the waves reflected from the welds on the right side were analysed.\par
For the POD analysis, we generated two sets of data for comparison. One set comprised purely random noise, commonly employed in literature to include noise effects. Figure \ref{Figure 18a} demonstrates an example T(0,1) signal with random noise on both linear and logarithmic scales, showing a consistent noise level marked by the shaded grey area. The SNR of the random noise was approximately 65 dB, determined based on the random noise measurements in the representative experiment. The other dataset included both coherent and random noise, as shown in Fig. \ref{Figure 18b}, generated using the proposed realistic simulation model. This signal exhibits a behaviour similar to the real experimental signal depicted in Fig. \ref{Figure 1}. The coherent noise is proportional to the signal amplitude and decreases with wave propagation due to material damping and energy spread, whereas random noise remains constant.\par

As mentioned earlier, the inspection range is typically determined by the point where the Call DAC drops below the Detection Threshold, following current industrial practice. In Fig. \ref{Figure 18}, the noise area was subjectively marked based on intuitive noise levels, with the Detection Threshold set at 6 dB above the noise level. Consequently, in both cases depicted in Fig. \ref{Figure 18}, the estimated inspection range is approximately 6.6 m. The estimation of the inspection range for these two cases, in Fig.\ref{Figure 18a} and Fig.\ref{Figure 18b}, shows similarity, as the intersection points fall within the range dominated by random noise, which exhibits similar levels for both cases. However, this method of determining inspection ranges presents two issues: firstly, the Detection Threshold is established based on subjective judgment of noise levels by the operator, and secondly, there lacks a quantified measure of confidence within the inspection range, which is not guaranteed to be 100\%. Hence, a more objective, automatic, and quantitative estimation procedure will be presented in this section based on POD analysis.\par

In practical guided wave monitoring, a near-field zone, as shown in Fig. \ref{Figure 1}, exists on either side of the transducer ring, characterized by artificially reduced amplitudes unsuitable for defect quantification\cite{e07_committee_practice_nodate}. In the simulation, this region was set at 0.8 m away from the transducer ring for illustrative purposes, as indicated in grey on the left side of Fig. \ref{Figure 18a} and Fig. \ref{Figure 18b}. Consequently, the estimation area was defined as 1.5 m to 7.4 m for sampling defect locations, and POD curves were computed along the pipe axis at intervals of 60 mm, approximately one wavelength of the T(0,1) wave. A total of 100 locations along the pipe axis were considered to calculate the POD curves for estimating the inspection range. \par
To quantify the defect response, DAC curves were determined based on the weld reflections, as shown in Fig. \ref{Figure 18}. In this study, we used the Flange DAC curve to quantify the reflection from a defect in terms of CSC. As indicated in Eq. (\ref{Equation 23-POD}), another key factor that determines the calculation of POD and PFA values is the Detection Threshold, \(\hat{a}_{0}\). In this study, \(\hat{a}_{0}\) is set to a value corresponding to 1\% PFA\cite{bayoumi2021approaches,rebillat2018peaks} to ensure fair comparisons across different cases. Due to variations in the noise level across cases, the threshold was adjusted accordingly to maintain a consistent PFA level.\par

\subsection{Generation of POD curves}
\begin{figure}
\centering 
\subfloat[]{\label{Figure 20a}\includegraphics{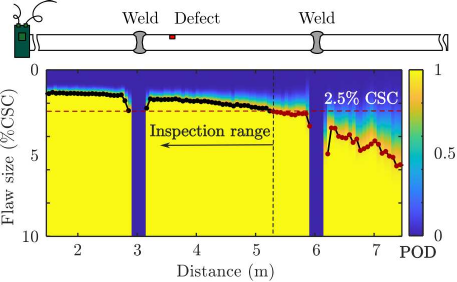}} \\
\subfloat[]{\label{Figure 20b}\includegraphics{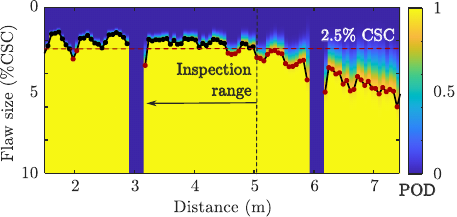}} 
\caption{POD maps calculated from two distinct datasets for the quantification of the inspection range. (a) POD map calculated based on the data with only random noise, (b) POD map calculated based on the data with both coherent noise and random noise. The POD map was created by calculating the POD curve at different distances from the transducer ring. The plotted line with dot markers represents the \(a_{90/95}\) flaw sizes at each distance. The black markers indicate values below the 2.5\% CSC threshold, while the red markers indicate values above it.}
\label{Figure 20}
\end{figure}

Figure \ref{Figure 19a} and Fig. \ref{Figure 19b} illustrate the linear regression relationship between the signal response and flaw size in terms of CSC when the defect is positioned at a distance of 3.54 m from the transducer ring. Figure \ref{Figure 19a} is based on the data set with only random noise, whereas Fig. \ref{Figure 19b} is based on the dataset with both coherent noise and random noise. At the chosen location, coherent noise dominates in Fig. \ref{Figure 19b}.
The estimates of \(\beta_{0}\) and \(\beta_{1}\), along with the corresponding standard error (SE) of the estimates, are illustrated in Fig. \ref{Figure 20a} and Fig. \ref{Figure 20b}. 
These coefficients were subsequently used in accordance with Eq. (\ref{Equation 23-POD}) to compute the POD curves. The notably high coefficient of determination (\(R^2\)) in both cases indicates a linear relationship between the signal response and the increasing flaw size in terms of CSC, aligning with the findings reported in\cite{demma2003reflection}.\par
The POD curves and their corresponding 95\% confidence intervals, calculated from two distinct datasets, are compared in Fig. \ref{Figure 19c}. It is evident that the POD curve, when including coherent noise, shifts to the right of the curve derived from data considering only random noise. This shift implies that, for the same-sized defect, a lower POD value is expected when the influence of coherent noise is considered. In POD analysis, the \(a_{90/95}\) value is typically of interest for comparison, representing the flaw size at which the system has 90\% POD with 95\% confidence. The \(a_{90/95}\) values determined for these two cases are 2.36\% CSC and 3.16\% CSC, respectively, indicating the risk of overestimating POD when only random noise is considered in the calculation of POD curves. As mentioned before, coherent noise is proportional to signal amplitude and decreases with distance, whereas random noise remains constant, becoming more important as the distance increases. Therefore, modelling coherent noise in the range where it dominates is necessary for a more precise estimation of the inspection range.

\subsection{Estimation of monitoring range}

For the quantitative estimation of the inspection range of a guided wave system for pipe monitoring, POD maps along the pipe axis were generated based on two distinct datasets. Figure \ref{Figure 20a} presents the POD map calculated based on the data with only random noise, where the horizontal axis corresponds to the distance of the defect away from the transducer ring, the vertical axis corresponds to the flaw sizes, and the colour-coded map displays the corresponding POD values. Figure \ref{Figure 20b} presents the POD map calculated based on the data with both coherent noise and random noise. In both figures, the \(a_{90/95}\) values at different distances are shown for quantitative comparison. In both POD maps, the \(a_{90/95}\) flaw size increases with the increasing distance away from the transducer ring, indicating a decay in the system performance with increasing detection distance. It is noted that the signal response and the calculation of the POD curves at the weld location differ from other locations due to the overlapping of defect reflections and weld reflections. Hence, the POD map at the weld locations is not presented here for a fair comparison.\par
In Fig. \ref{Figure 20a}, the \(a_{90/95}\) curve exhibits different characteristics in the area near the transducer ring compared to that farther from the transducer ring. In the area near the transducer ring, the defect reflection generally exhibits a higher amplitude, resulting in a higher SNR with a constant random noise level. The signal response calculated based on the Flange DAC curve is less affected by random noise, leading to a smooth change in the \(a_{90/95}\) curve in the nearby area. Conversely, for defects located far from the transducer ring, the signal response is significantly influenced by random noise, resulting in larger variations in the \(a_{90/95}\) values.\par
In Fig. \ref{Figure 20b}, the effect of coherent noise is included. The \(a_{90/95}\) curve demonstrates evident fluctuations in the region close to the transducer ring, attributed to the significant influence of coherent noise on the defect reflection. In the region far from the transducer ring, random noise dominates the noise signal. The POD map displays variations similar to those observed in the POD map calculated based on the dataset with only random noise in Fig. \ref{Figure 20a}.\par
For determining the inspection range with a specific sensitivity requirement, for example, 2.5\% CSC as marked in Fig. \ref{Figure 20}, a threshold line can be drawn, and the furthest distance at which the \(a_{90/95}\) curve permanently falls below this threshold line specifies the inspection range. The locations where the \(a_{90/95}\) flaw size exceeds 2.5\% CSC are highlighted by the red points in Fig. \ref{Figure 20a} and Fig. \ref{Figure 20b}. For the same required \(a_{90/95}\) flaw size, Fig. \ref{Figure 20b} indicates that there are short regions with a lower POD within the inspection range due to the presence of coherent noise. The inspection range determined based on the dataset with only random noise data is approximately 5.28 m, while that determined based on the dataset with both coherent noise and random noise is about 5.04 m, albeit with some locations exhibiting lower POD values.\par
In comparison with the current industry practice for determining the inspection range, which stands at about 6.6 m for both cases based on the intersection between the Call DAC and the Detection Threshold as depicted in Fig. \ref{Figure 18}, the method proposed in this study based on the POD map calculated using the simulation-generated realistic guided wave signals provides a more detailed quantitative estimation of the system's detection capabilities within the inspection range.\par

\section{Conclusion}\label{Conclusion}
In this paper, we propose a numerical framework for generating realistic guided wave signals for the purpose of reliability estimation. Through a comprehensive analysis of experimental signals, we investigate the characteristics and composition of noise, including coherent and random noise. We find that random noise can be effectively modelled as band-limited Gaussian white noise. Moreover, we demonstrate that coherent noise can be modelled based on the transducer transfer functions, derived from a comparison between the representative experimental signals and the idealised calibration simulation signals. Through both visual observation and statistical analysis, we demonstrate that the proposed framework successfully generates realistic guided wave signals that accurately represent the noise characteristics found in actual guided wave signals. \par
An application of the realistic simulation-generated signals was then illustrated. The reliability of the guided wave system for pipe monitoring used in this study was assessed in terms of the inspection range via POD analysis. A comparison of the POD curves generated using datasets containing only random noise and those with both coherent and random noise highlights the crucial importance of considering coherent noise for more accurate reliability estimation in the range where coherent noise is dominating. Additionally, compared to the current industry practice for inspection range determination, the POD analysis using realistic simulation-generated data offers more detailed quantitative insights.\par
The numerical framework proposed in this study provides, for the first time, an efficient method for generating realistic guided wave signals tailored for practical applications. This framework facilitates the performance evaluation of a guided wave monitoring system under diverse and realistic conditions within the SHM context, offering enhanced flexibility and feasibility. While the proposed numerical framework was illustrated based on the generation of realistic guided wave signals for pipe monitoring, it can be applied to general ultrasonic testing scenarios and other applications where the measurement uncertainty needs to be faithfully represented in the simulation. For example, realistic bulk wave signals can be generated for phased array ultrasonic testing applications by incorporating representations of both random and coherent noise originating from the measurement system. This further enables the production of realistic phased array images tailored for a range of diagnostic purposes. Furthermore, our framework exhibits potential as a versatile tool for generating realistic simulation data with broader applications, including training machine learning models. This capability addresses challenges related to data scarcity in the NDT and SHM communities.

\section*{Acknowledgement}

The authors would like to thank Dr Thomas Vogt and Dr Stefano Mariani for the valuable discussions with them, thank Prof Peter Cawley, Prof Michael Lowe, and Prof Constantinos Soutis for manuscript reviewing, and thank Antonio De Sanctis for his help with the experimental work in this article.

\bibliographystyle{IEEEtran}
\bibliography{Mybib}

\begin{thebibliography}{10}
\providecommand{\url}[1]{#1}
\csname url@samestyle\endcsname
\providecommand{\newblock}{\relax}
\providecommand{\bibinfo}[2]{#2}
\providecommand{\BIBentrySTDinterwordspacing}{\spaceskip=0pt\relax}
\providecommand{\BIBentryALTinterwordstretchfactor}{4}
\providecommand{\BIBentryALTinterwordspacing}{\spaceskip=\fontdimen2\font plus
\BIBentryALTinterwordstretchfactor\fontdimen3\font minus \fontdimen4\font\relax}
\providecommand{\BIBforeignlanguage}[2]{{%
\expandafter\ifx\csname l@#1\endcsname\relax
\typeout{** WARNING: IEEEtran.bst: No hyphenation pattern has been}%
\typeout{** loaded for the language `#1'. Using the pattern for}%
\typeout{** the default language instead.}%
\else
\language=\csname l@#1\endcsname
\fi
#2}}
\providecommand{\BIBdecl}{\relax}
\BIBdecl

\bibitem{cawley2003practical}
P.~Cawley, ``Practical long range guided wave inspection—managing complexity,'' in \emph{AIP Conference Proceedings}, vol. 657, no.~1.\hskip 1em plus 0.5em minus 0.4em\relax American Institute of Physics, 2003, pp. 22--40.

\bibitem{Peter10.1115/1.2789107}
\BIBentryALTinterwordspacing
M.~J.~S. Lowe, D.~N. Alleyne, and P.~Cawley, ``{The Mode Conversion of a Guided Wave by a Part-Circumferential Notch in a Pipe},'' \emph{Journal of Applied Mechanics}, vol.~65, no.~3, pp. 649--656, 09 1998. [Online]. Available: \url{https://doi.org/10.1115/1.2789107}
\BIBentrySTDinterwordspacing

\bibitem{croxford2007strategies}
A.~J. Croxford, P.~D. Wilcox, B.~W. Drinkwater, and G.~Konstantinidis, ``Strategies for guided-wave structural health monitoring,'' \emph{Proceedings of the Royal Society A: Mathematical, Physical and Engineering Sciences}, vol. 463, no. 2087, pp. 2961--2981, 2007.

\bibitem{cawley2012guided}
P.~Cawley, F.~Cegla, and A.~Galvagni, ``Guided waves for ndt and permanently-installed monitoring,'' \emph{Insight-Non-Destructive Testing and Condition Monitoring}, vol.~54, no.~11, pp. 594--601, 2012.

\bibitem{Mariani:2023}
S.~Mariani, T.~vogt, S.~Heinlein, and P.~Cawley, ``The performance of a guided wave pipe monitoring system over extended periods of field operation,'' \emph{Materials Evaluation}, 2023.

\bibitem{giurgiutiu2010guided}
V.~Giurgiutiu and C.~Soutis, ``Guided wave methods for structural health monitoring,'' \emph{Encyclopedia of aerospace engineering}, 2010.

\bibitem{Vogt:2021}
\BIBentryALTinterwordspacing
T.~Vogt, S.~Heinlein, J.~Milewczyk, S.~Mariani, R.~Jones, and P.~Cawley, \emph{Guided Wave Monitoring of Industrial Pipework - Improved Sensitivity System and Field Experience}, 2021, pp. 819--829. [Online]. Available: \url{http://dx.doi.org/10.1007/978-3-030-64594-6\_79}
\BIBentrySTDinterwordspacing

\bibitem{Cawley:2019}
\BIBentryALTinterwordspacing
P.~Cawley, ``Ultrasonic structural health monitoring - current applications and potential.''\hskip 1em plus 0.5em minus 0.4em\relax IEEE, 2019, pp. 2107--2109. [Online]. Available: \url{http://dx.doi.org/10.1109/ultsym.2019.8926176}
\BIBentrySTDinterwordspacing

\bibitem{cawley2021development}
{P. Cawley}, ``A development strategy for structural health monitoring applications,'' \emph{Journal of Nondestructive Evaluation, Diagnostics and Prognostics of Engineering Systems}, vol.~4, no.~4, p. 041012, 2021.

\bibitem{mesnil2017guided}
O.~Mesnil, B.~Chapuis, and T.~Druet, ``Guided waves for structural health monitoring,'' in \emph{56th Annual Conference of the British Institute of Non-Destructive Testing, NDT 2017}.\hskip 1em plus 0.5em minus 0.4em\relax British Institute of Non-Destructive Testing, 2017.

\bibitem{hdbk2009nondestructive}
``Nondestructive evaluation system reliability assessment,'' Department of Defense Handbook, Standard, 2009.

\bibitem{ModelAssistedPODWorkingGroup}
M.-A. P.~W. Group, ``Model-assisted pod working group,'' https://static.cnde.iastate.edu/mapod/About\%20Us.htm.

\bibitem{thompson2009recent}
R.~B. Thompson, L.~J. Brasche, E.~Lindgren, P.~Swindell, and W.~P. Winfree, ``Recent advances in model-assisted probability of detection,'' in \emph{4th European-American workshop on reliability of NDE}, no. LF99-9094, 2009.

\bibitem{du2019multifidelity}
X.~Du and L.~Leifsson, ``Multifidelity model-assisted probability of detection via cokriging,'' \emph{NDT \& E International}, vol. 108, p. 102156, 2019.

\bibitem{yilmaz2022model}
B.~Yilmaz, D.~Smagulova, and E.~Jasiuniene, ``Model-assisted reliability assessment for adhesive bonding quality evaluation with ultrasonic ndt,'' \emph{NDT \& E International}, vol. 126, p. 102596, 2022.

\bibitem{Pierre2023}
\BIBentryALTinterwordspacing
P.~Calmon, S.~Sharma, O.~Mesnil, and B.~Chapuis, ``{Experimental validation of MAPOD methodology for SHM applied to the detection of growing cracks in a metallic part (Conference Presentation)},'' in \emph{8th International Workshop on Reliability of NDT/NDE}, D.~Kanzler and N.~G. Meyendorf, Eds., vol. PC12491, International Society for Optics and Photonics.\hskip 1em plus 0.5em minus 0.4em\relax SPIE, 2023, p. PC1249105. [Online]. Available: \url{https://doi.org/10.1117/12.2662747}
\BIBentrySTDinterwordspacing

\bibitem{falcetelli2023model}
F.~Falcetelli, N.~Yue, L.~Rossi, G.~Bolognini, F.~Bastianini, D.~Zarouchas, and R.~Di~Sante, ``A model-assisted probability of detection framework for optical fiber sensors,'' \emph{Sensors}, vol.~23, no.~10, p. 4813, 2023.

\bibitem{aldrin2009model}
J.~C. Aldrin, J.~S. Knopp, E.~A. Lindgren, and K.~V. Jata, ``Model-assisted probability of detection evaluation for eddy current inspection of fastener sites,'' in \emph{AIP Conference Proceedings}, vol. 1096, no.~1.\hskip 1em plus 0.5em minus 0.4em\relax American Institute of Physics, 2009, pp. 1784--1791.

\bibitem{smith2007model}
K.~Smith, B.~Thompson, B.~Meeker, T.~Gray, and L.~Brasche, ``Model-assisted probability of detection validation for immersion ultrasonic application,'' in \emph{AIP Conference Proceedings}, vol. 894, no.~1.\hskip 1em plus 0.5em minus 0.4em\relax American Institute of Physics, 2007, pp. 1816--1822.

\bibitem{moriot2018model}
J.~Moriot, N.~Quaegebeur, A.~Le~Duff, and P.~Masson, ``A model-based approach for statistical assessment of detection and localization performance of guided wave--based imaging techniques,'' \emph{Structural Health Monitoring}, vol.~17, no.~6, pp. 1460--1472, 2018.

\bibitem{howard2017probability}
R.~Howard and F.~Cegla, ``On the probability of detecting wall thinning defects with dispersive circumferential guided waves,'' \emph{NDT \& E International}, vol.~86, pp. 73--82, 2017.

\bibitem{howard2017detectability}
{R. Howard and F. Cegla}, ``Detectability of corrosion damage with circumferential guided waves in reflection and transmission,'' \emph{NDT \& E International}, vol.~91, pp. 108--119, 2017.

\bibitem{foucher2018new}
F.~Foucher, R.~Fernandez, S.~Leberre, and P.~Calmon, ``New tools in civa for model assisted probability of detection (mapod) to support nde reliability studies,'' in \emph{NDE of Aerospace Materials \& Structures 2018}, 2018, pp. 32--43.

\bibitem{khurjekar2019accounting}
I.~D. Khurjekar and J.~B. Harley, ``Accounting for physics uncertainty in ultrasonic wave propagation using deep learning,'' \emph{arXiv preprint arXiv:1911.02743}, 2019.

\bibitem{mariani2020compensation}
S.~Mariani, S.~Heinlein, and P.~Cawley, ``Compensation for temperature-dependent phase and velocity of guided wave signals in baseline subtraction for structural health monitoring,'' \emph{Structural Health Monitoring}, vol.~19, no.~1, pp. 26--47, 2020.

\bibitem{herdovics2019compensation}
B.~Herdovics and F.~Cegla, ``Compensation of phase response changes in ultrasonic transducers caused by temperature variations,'' \emph{Structural Health Monitoring}, vol.~18, no.~2, pp. 508--523, 2019.

\bibitem{liu2017efficient}
C.~Liu, J.~Dobson, and P.~Cawley, ``Efficient generation of receiver operating characteristics for the evaluation of damage detection in practical structural health monitoring applications,'' \emph{Proceedings of the Royal Society A: Mathematical, Physical and Engineering Sciences}, vol. 473, no. 2199, p. 20160736, 2017.

\bibitem{luleci2022generative}
F.~Luleci, F.~N. Catbas, and O.~Avci, ``Generative adversarial networks for data generation in structural health monitoring,'' \emph{Frontiers in Built Environment}, vol.~8, p. 816644, 2022.

\bibitem{e07_committee_practice_nodate}
\BIBentryALTinterwordspacing
{E07 Committee}, ``Practice for guided wave testing of above ground steel pipework using piezoelectric effect transduction.'' [Online]. Available: \url{https://www.astm.org/e2775-16.html}
\BIBentrySTDinterwordspacing

\bibitem{vogt2022multiple}
T.~Vogt, B.~Pavlakovic, and P.~Cawley, ``A multiple-echo calibration technique for guided wave testing,'' \emph{Materials Evaluation}, vol.~80, no.~6, pp. 32--43, 2022.

\bibitem{evans2010reliability}
T.~Vogt and M.~Evans, ``Reliability of guided wave testing.''\hskip 1em plus 0.5em minus 0.4em\relax 4th European-American Workshop on Reliability of NDE, 2010.

\bibitem{mariani2019location}
S.~Mariani, S.~Heinlein, and P.~Cawley, ``Location specific temperature compensation of guided wave signals in structural health monitoring,'' \emph{IEEE Transactions on Ultrasonics, Ferroelectrics, and Frequency Control}, vol.~67, no.~1, pp. 146--157, 2019.

\bibitem{Material}
MakeItFrom, ``Makeitfrom,'' https://www.makeitfrom.com/material-properties/EN-1.0425-P265GH-Non-Alloy-Steel.

\bibitem{pavlakovic1997disperse}
B.~Pavlakovic, M.~Lowe, D.~Alleyne, and P.~Cawley, ``Disperse: A general purpose program for creating dispersion curves,'' \emph{Review of Progress in Quantitative Nondestructive Evaluation: Volume 16A}, pp. 185--192, 1997.

\bibitem{wilcox2003omni}
P.~D. Wilcox, ``Omni-directional guided wave transducer arrays for the rapid inspection of large areas of plate structures,'' \emph{IEEE transactions on ultrasonics, ferroelectrics, and frequency control}, vol.~50, no.~6, pp. 699--709, 2003.

\bibitem{Davies10.1063/1.2184522}
\BIBentryALTinterwordspacing
J.~Davies, F.~Simonetti, M.~Lowe, and P.~Cawley, ``{Review of Synthetically Focused Guided Wave Imaging Techniques With Application to Defect Sizing},'' \emph{AIP Conference Proceedings}, vol. 820, no.~1, pp. 142--149, 03 2006. [Online]. Available: \url{https://doi.org/10.1063/1.2184522}
\BIBentrySTDinterwordspacing

\bibitem{birnie2016analysis}
C.~Birnie, K.~Chambers, D.~Angus, and A.~L. Stork, ``Analysis and models of pre-injection surface seismic array noise recorded at the aquistore carbon storage site,'' \emph{Geophysical Journal International}, vol. 206, no.~2, pp. 1246--1260, 2016.

\bibitem{martinez2015computational}
W.~L. Martinez and A.~R. Martinez, \emph{Computational statistics handbook with MATLAB}.\hskip 1em plus 0.5em minus 0.4em\relax CRC press, 2015, vol.~22.

\bibitem{hayashi2005mode}
T.~Hayashi and M.~Murase, ``Mode extraction technique for guided waves in a pipe,'' \emph{Nondestructive testing and evaluation}, vol.~20, no.~1, pp. 63--75, 2005.

\bibitem{Mike10.1115/1.2789107}
\BIBentryALTinterwordspacing
M.~J.~S. Lowe, D.~N. Alleyne, and P.~Cawley, ``{The Mode Conversion of a Guided Wave by a Part-Circumferential Notch in a Pipe},'' \emph{Journal of Applied Mechanics}, vol.~65, no.~3, pp. 649--656, 09 1998. [Online]. Available: \url{https://doi.org/10.1115/1.2789107}
\BIBentrySTDinterwordspacing

\bibitem{sanchez2006characterization}
A.~L.~L. S{\'a}nchez and L.~W. Schmerr~Jr, ``Characterization of an ultrasonic nondestructive measurement system,'' 2006.

\bibitem{lopez2005ultrasonic}
A.~Lopez-Sanchez, ``Ultrasonic system models and measurements,'' Ph.D. dissertation, Iowa State University, 2005.

\bibitem{wiener1964wiener}
N.~Wiener, ``The wiener rms (root mean square) error criterion in filter design and prediction,'' 1964.

\bibitem{cicero2009potential}
T.~Cicero, P.~Cawley, F.~Simonetti, and S.~Rokhlin, ``Potential and limitations of a deconvolution approach for guided wave structural health monitoring,'' \emph{Structural Health Monitoring}, vol.~8, no.~5, pp. 381--395, 2009.

\bibitem{huthwaite2014accelerated}
P.~Huthwaite, ``Accelerated finite element elastodynamic simulations using the gpu,'' \emph{Journal of Computational Physics}, vol. 257, pp. 687--707, 2014.

\bibitem{KARA2010998}
\BIBentryALTinterwordspacing
F.~Kara, J.~Navarro, and R.~L. Allwood, ``Effect of thickness variation on collapse pressure of seamless pipes,'' \emph{Ocean Engineering}, vol.~37, no.~11, pp. 998--1006, 2010. [Online]. Available: \url{https://www.sciencedirect.com/science/article/pii/S0029801810000788}
\BIBentrySTDinterwordspacing

\bibitem{pyshmintsev2016evolution}
I.~Pyshmintsev, I.~Veselov, A.~Yakovleva, M.~Lobanov, and S.~Danilov, ``Evolution of the texture of low-carbon microalloyed pipe steel in the seamless pipe manufacturing process,'' in \emph{AIP Conference Proceedings}, vol. 1785, no.~1.\hskip 1em plus 0.5em minus 0.4em\relax AIP Publishing, 2016.

\bibitem{malachowski2014investigation}
J.~Malachowski, V.~Hutsaylyuk, P.~Yukhumets, R.~Dmitryenko, G.~Beliaiev, and I.~Prudkii, ``Investigation of the stress-strain state of seamless pipe in the initial state,'' \emph{Archive of mechanical engineering}, vol.~61, no.~4, pp. 595--607, 2014.

\bibitem{gandossi2010probability}
L.~Gandossi and C.~Annis, ``Probability of detection curves: Statistical best-practices,'' \emph{ENIQ report}, vol.~41, 2010.

\bibitem{bayoumi2021approaches}
A.~B. Bayoumi, I.~Mueller, T.~Vogt, and P.~Kraemer, ``Approaches combining multiple paths to establish the probability of detection of a guided wave-based structural health monitoring system,'' \emph{Prague, Czech Republic, European NDT CM}, 2021.

\bibitem{rebillat2018peaks}
M.~R{\'e}billat, O.~Hmad, F.~Kadri, and N.~Mechbal, ``Peaks over threshold--based detector design for structural health monitoring: Application to aerospace structures,'' \emph{Structural Health Monitoring}, vol.~17, no.~1, pp. 91--107, 2018.

\bibitem{demma2003reflection}
A.~Demma, P.~Cawley, M.~Lowe, and A.~Roosenbrand, ``The reflection of the fundamental torsional mode from cracks and notches in pipes,'' \emph{The Journal of the Acoustical Society of America}, vol. 114, no.~2, pp. 611--625, 2003.

\end{thebibliography}

\end{document}